\documentclass[10pt,journal,compsoc]{IEEEtran}

\usepackage{subfigure}
\usepackage{graphicx}
\usepackage{epsfig}
\usepackage{url}
\usepackage{amsfonts}
\usepackage{amssymb}
\usepackage{mdwmath} 
\usepackage{mdwtab}
\usepackage{cite}
\usepackage{booktabs}
\usepackage{tabularx}
\usepackage{multirow} 
\usepackage{caption}
\usepackage[font=small,labelfont=sf,tableposition=top, margin=10pt]{caption}
\usepackage{amsmath}
\usepackage{algorithm} 
\usepackage{algorithmic}
\usepackage{color}
\usepackage{verbatim}

\floatname{algorithm}{Procedure}

\begin{document}

\title{Sparsity based Efficient Cross-Correlation Techniques in Sensor Networks}

\author{Prasant~Misra,~\IEEEmembership{Senior~Member,~IEEE,}
       Wen~Hu,~\IEEEmembership{Senior~Member,~IEEE,}
       Mingrui~Yang,~\IEEEmembership{Member,~IEEE,}
			 Marco~Duarte,~\IEEEmembership{Senior~Member,~IEEE,}
 			 Sanjay~Jha,~\IEEEmembership{Senior~Member,~IEEE}
\IEEEcompsocitemizethanks{\IEEEcompsocthanksitem P. Misra is with the TCS Innovation Labs, Bangalore, India. E-mail: prasant.misra@tcs.com\protect\\
\IEEEcompsocthanksitem M. Yang is at the Case Western Reserve University, USA. E-mail: mingrui.yang@case.edu\protect\\
\IEEEcompsocthanksitem M. Duarte is at the University of Massachusetts, Amherst, USA. E-mail: mduarte@ecs.umass.edu\protect\\
\IEEEcompsocthanksitem W. Hu and S. Jha are at the University of New South Wales, Sydney, Australia. E-mail: \{wen.hu, sanjay\}@cse.unsw.edu.au
}}


\IEEEtitleabstractindextext{%
\begin{abstract}
Cross-correlation is a popular signal processing technique used in numerous location tracking systems for obtaining reliable \emph{range} information.
However, its efficient design and practical implementation has not yet been achieved on mote platforms that are typical in wireless sensor network due to resource constrains.
In this paper, we propose \emph{SparseS-XCorr}: cross-correlation via structured sparse representation, a new computing framework for ranging based on $\ell^1$- minimization \cite{donoho:l1} and structured sparsity.
The key idea is to compress the ranging signal samples on the mote by efficient random projections and transfer them to a central device; where a convex optimization process estimates the range by exploiting the sparse signal structure in the proposed \textit{correlation} dictionary. 
Through theoretical validation, extensive empirical studies and experiments on an \textit{end-to-end} acoustic ranging system implemented on resource limited off-the-shelf sensor nodes, we show that the proposed framework can achieve up to \emph{two orders of magnitude} better performance compared to other approaches such as working on DCT domain and downsampling.
Compared to the standard cross-correlation, it is able to obtain range estimates with a bias of $2$-$6$\,cm with $30$\% and approximately $100$\,cm with $5$\% compressed measurements.
Its structured sparsity model is able to improve the ranging accuracy by $40$\% under challenging recovery conditions (such as high compression factor and low signal-to-noise ratio) by overcoming limitations due to dictionary coherence.
\end{abstract}

}

\maketitle
\IEEEdisplaynontitleabstractindextext
\IEEEpeerreviewmaketitle

\section{Introduction} \label{sec:intro}
Location sensing is a \emph{vital} enabling technology for numerous applications in the field of binaural science, acoustic source detection, target motion analysis, sensor networking, mobile robot navigation, mobile computing, etc.
While GPS remains to be the \emph{de facto} solution for outdoor positioning, its limitation to service GPS denied environments (such as indoor and obstructed outdoor) makes location estimation - \emph{still} - a fundamental problem.
Localization is a \emph{two} step process.
The \emph{first} step is to measure the separation distance (or range) of the unknown entity (that needs to be localized) from at least three positioned entities (or known locations).
These measurements are subsequently utilized in the \emph{second} step that multilaterates the 
position estimate using a spatially constrained optimization framework.
This \emph{strong} dependency of the reliability of positioning accuracy on the distance measurement makes ranging a \emph{crucial} prerequisite for localization.
\vspace{2mm}
\newline
\noindent
\textbf{Challenges.} Acoustic and radio ranging technologies have matured significantly in the last few decades. 
It is now well understood that highly accurate results can be achieved by measuring the \emph{travel time}\footnote{Travel time is interchangeably referred to as: time-of-flight (ToF), time-of-arrival (ToA), propagation delay, or time delay in the rangefinding literature.} 
of the ranging signal. 
However, the resources required for signal detection are a deciding factor for the cost, size and weight of the sensing platform; and this essentially strikes a \emph{trade-off} between (localization accuracy, coverage range) and energy efficiency.
Low-cost and low-power systems estimate the arrival time of the pulse by utilizing simple detection schemes (such as empirical thresholding of the leading pulse edge\cite{Priyanthathesis2005:Cricket}). 
Nevertheless, they turn out to be less reliable due to their limited computational capability to counter environmental noise and multipath reflections \cite{Misra2011:coverage}.
An established methodology to overcome these limitations is to \emph{broaden} the range of signal frequencies and distribute the energy between the various multiple paths; and subsequently apply a \emph{matched filter} at the receiver end to count the elapsed time samples by resolving those multiple propagation paths.
Its benefits are \emph{two} fold as broadband signals reduce the chance of the entire signal fading at any particular time, while matched filters allow for their processing and form a strong pulse at the line-of-sight (LoS) path by increasing the overall signal-to-noise ratio (SNR) without using excess transmission power.
\newline
\indent
There are numerous in-air and underwater ranging systems\cite{Hazas2006:USBroadband,Kushwaha2005,Girod2006:AENSBox,Chunyi2007:BeepBeep,Misra2011:tweet} that have widely used these techniques to deliver remarkable (accuracy vs. range) performance, but at the expense of specialized computing platforms (such as DSP processors) that are both power intensive and costly.
Such stringent needs pose a major challenge to the field of wireless sensor networks (WSN) that aim to achieve similar functional capability on constrained devices with high restrictions on data sensing rates, link bandwidth, computational speed, battery life and memory capacity (less than $50$\,kB of code memory and $10$\,kB RAM)\cite{whitehouse2002:calamari,Sallai04acousticranging}.
This has been the primal factor that has greatly limited the realisation of sophisticated algorithms (such as the matched filter).
This problem can be simplified by designing a light-weight signal detection and post-processing mechanism that not only serves the purpose of sample counting, but is also suitable for running on constrained embedded platforms typically used in WSNs.
Motivated by the need to design such a mechanism, we propose \emph{Struct-Sparse-XCorr}.
\vspace{2mm}
\newline
\noindent
\textbf{Contributions.} 
\emph{Struct-Sparse-XCorr (or StructS-XCorr)}: cross-correlation via structured sparse representation is a new computing framework for ranging based on $\ell^1$\,-\,minimization\cite{donoho:l1} and structured sparsity.
It is based on a mechanism to \emph{compress} and transmit the condensed ranging data to a more resourceful offloading device (or base-station), wherein the time delay of the ranging signal can be efficiently recovered to determine the range.
Cross-correlation is the \emph{conventional} method of obtaining this parameter; but, given its sparse information content and structure, we make use of the theoretical results in structured sparse approximation to achieve a similar performance. 
The underlying information theory suggests that a signal can be recovered by $\ell^1$\,-\,minimization\cite{donoho:l1}, when its representation is sufficiently \emph{sparse} with respect to an over-complete \emph{dictionary} of base elements.
The recovery model (or the optimization framework that bear resemblance to Lasso in statistics \cite{Zhao2006:Lasso1,Tibshirani1996:Lasso2}), instead of penalizing the number of nonzero coefficients directly (e.g., $\ell^0$-norm)\cite{Amaldi1998:l0}, penalizes the $\ell^1$\,-\,norm of the sparse coefficients in the linear combination.
\newline
\indent
We propose a \emph{new} dictionary that combines the information sparsity along the time-delay search dimension, and achieves up to \emph{two} order of magnitude better sparse representation and performance compared to standard approaches such as working on DCT domain and downsampling.
\emph{StructS-XCorr} overcomes ranging inaccuracies induced by dictionary coherence by approximately $40$\% for signals subjected to high compression factor and/or received with low SNR levels. 
\newline
\indent
We empirically validate our hypothesis in real-world indoor and outdoor setups.
With respect to cross-correlation, we show that \emph{StructS-XCorr} obtains range estimates with a relative error of less than $2$\,cm by using $30$\% compressed measurements, and approximately $60$\,cm relative error with $5$\% measurements only.
We also address the problems of \emph{slower} compression speed and \emph{incorrect} peak identification (important for estimating range) by devising a divide-and-conquer method.
\newline
\indent
We present the design and implementation of an \textit{end-to-end} acoustic ranging system consisting of Tmote Invent (receiver) nodes and a custom built audio (transmitter) node.
The results show a relative ranging and $2$D position error of less than $4$\,cm over cross-correlation using $30$-$40$\% compressed measurements, but with significant energy savings of an order of magnitude \emph{two}.
\newline
\indent
To support our contributions; we present the design of \emph{SparseS-Xcorr} and its emprirical studies in the next section, which is then followed by the description of the acoustic ranging system and its evaluation in Section~\ref{sec:evaluation}. 
Finally, we survey related work in Section~\ref{sec:related_work}, and summarize the paper with concluding remarks in Section~\ref{sec:conclusion}.

\section{The Design of StructS-XCorr}
To ground our discussion, in this section, we first present the details of time-based ranging using cross-correlation and structured sparse approximation.
We build on these learning to cast the ranging problem into the new computation framework of \emph{StructS-XCorr}, and then follow it up with its empirical analysis.

\subsection{An Overview of Time-based Ranging}

\begin{figure}[t]
\begin{center}
\subfigure[]{\label{fig:xcorr_a}\includegraphics[width=1.7in]{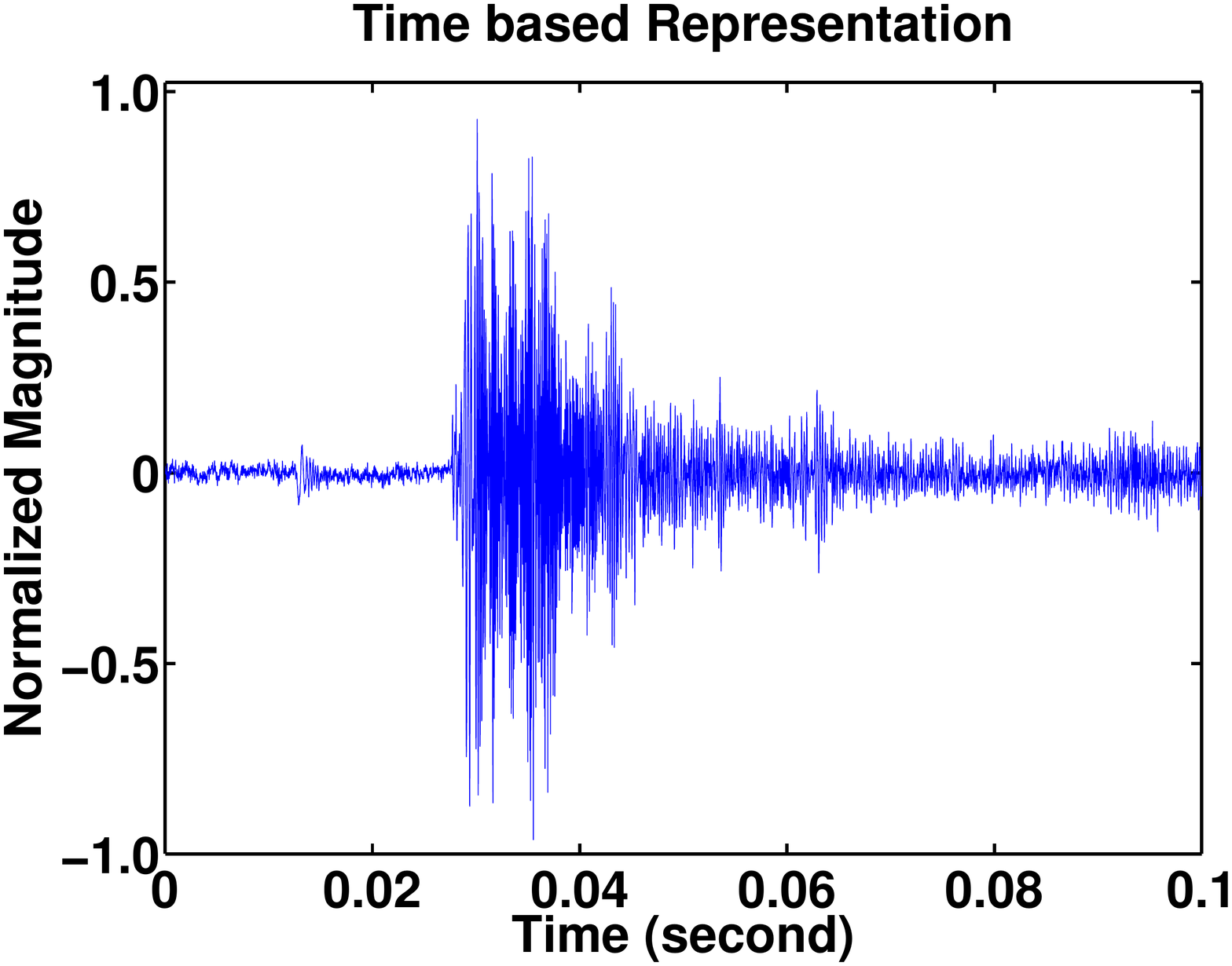}}
\subfigure[]{\label{fig:xcorr_b}\includegraphics[width=1.7in]{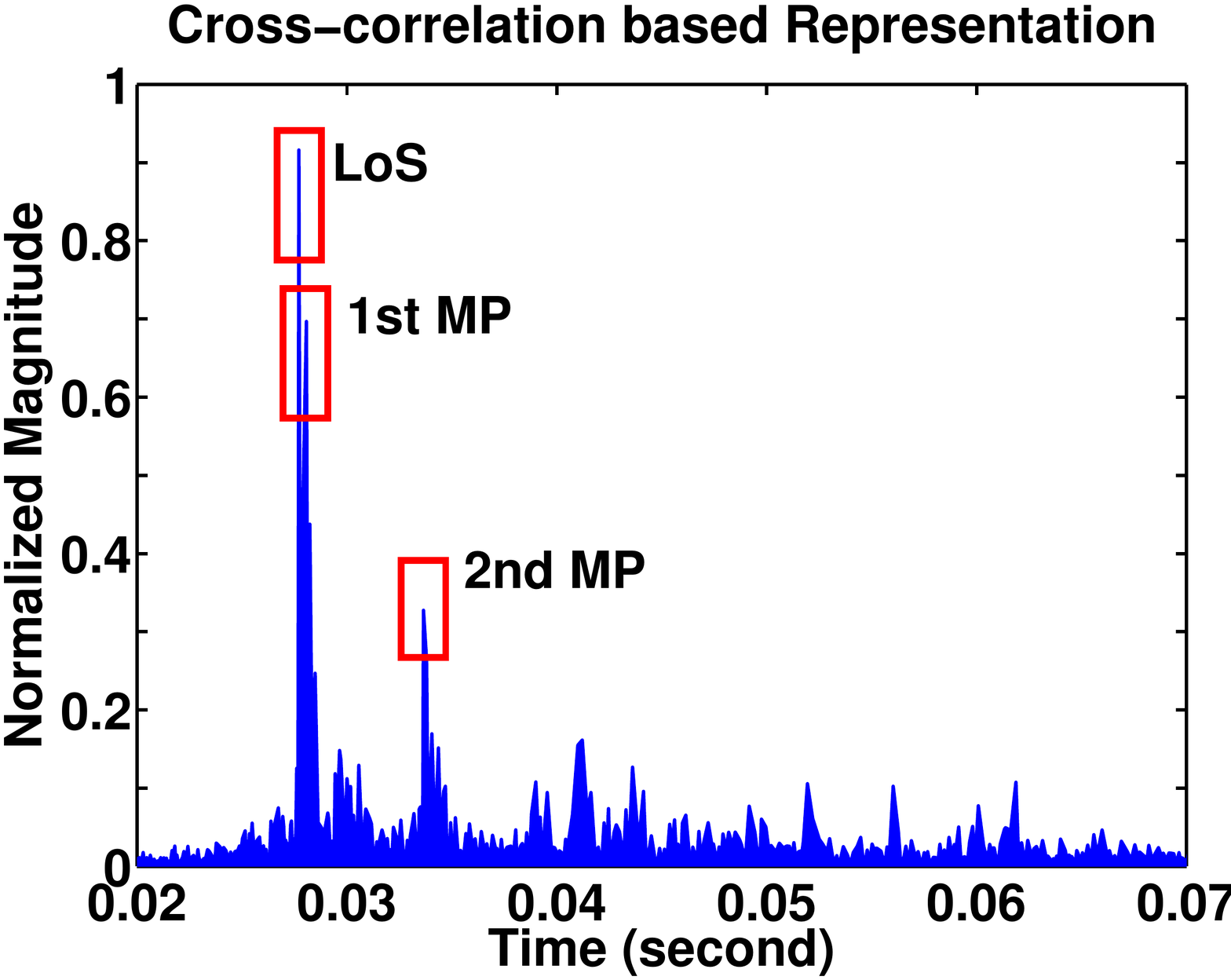}}
\end{center}
\vspace{-4mm}
\caption{\textbf{Range estimation by cross-correlation.} \emph{The information content is sparse as the time delay value corresponding to the correlation peak is only useful. It also depicts the same received waveform in two different (time and cross-correlation) representations. Note that in the figure: LOS stands for line-of-sight and MP is expanded as multipath.}} 
\vspace{-2mm}
\label{fig:xcorr}
\end{figure}

Previous studies have shown the most successful techniques for estimating the precise distance between two devices are based on measuring the \emph{travel time} of the signal propagation between them\cite{misra2015:acoustic_rangefinders}.
The reliability of this measurement depends on many factors, of which robustness of examining and estimating the energy of the received signal is one of them.
In this regard, \emph{matched filter} is the state-of-the-art in detection technology.
\newline
\indent
A matched filter is implemented by \emph{cross-correlating} the received signal $x(t)$ with the transmitted signal replica $p(t)$.
Cross-correlation (\emph{X-Corr}) of $p(t)$ and $x(t)$ is a sequence $s(\tau)$ defined~as:
\begin{equation}\label{eq:xcorr}
   s(\tau) = \int_{t=-\infty}^{t=+\infty} p(t+\tau)x(t) 
\end{equation}
where the index of $\tau \in \mathbb{R}$ is the time shift (or lag) parameter.
This operation \textbf{s($\tau)$} results in correlation peaks where the position of the peaks provides a measure of the arrival time of the different multipaths.
The index of the \emph{first} tallest correlation peak is the estimate of the pulse arrival time of the LOS path, which is a direct measure of the range.
Generally, $x(t)$ is acquired for a (finite) minimum time $t = t_{a}$ given by: 
\begin{equation}\label{eq:acqtime}
  t_{a} \geq \left(\frac{d_{c}}{v_{s}} + t_{p} + t_{r}\right)
\end{equation}
where, $d_{c}$ is the channel length between the transmitter and the receiver, $v_{s}$ is the speed of the ranging signal in the medium, $t_{p}$ is the time-period of the transmitted signal $p(t)$, and $t_{r}$ is the approximate reverberation time within which the echoes from the transmitted pulse should have fallen below an acceptable level before the next pulse is emitted.
The corresponding discrete-time signal of $p(t)$ and $x(t)$ obtained at a sampling rate (hertz/samples per second) of $F_{s}$ is given as: $p[n_{p}] = p[t_{p} F_{s}]$ and $x[n_{a}] = x[t_{a} F_{s}]$ $0 \leq n_{p},n_{a} \leq \infty$.
Therefore, $p(t)$ and $x(t)$ can be represented as vectors $\textbf{p} \in \mathbb{R}^{n_{p}}$ and $\textbf{x} \in \mathbb{R}^{n_{a}}$.
The time delay is obtained by finding:
\begin{equation}\label{eq:maxlikelihood}
\hat{\tau} = \arg \max_\tau |\textbf{s}(\tau)|^{2}.
\end{equation}
\noindent
\textbf{Road-map.} The computing operation of $\hat{\tau}$ (Eq.~\ref{eq:maxlikelihood}) is \emph{expensive}, and demands high memory, computation and energy resources.
Considering the constraints of typical WSN platforms, it is desirable
to scale down its complexity by a simpler process while still being capable of precisely estimating $\hat{\tau}$.
This motivates the scope for a \emph{new} framework.
\newline
\indent
Fig.~\ref{fig:xcorr}(a) shows a received signal trace recorded for a duration of $0.1$\,s sampled at $48$\,kHz, and its cross-correlation with the reference copy (a linear chirp of $1$-$20$\,kHz/$0.01$\,s) is depicted in Fig.~\ref{fig:xcorr}(b).
Ideally, only a \emph{single} dominant peak should be observed at the correct time shift; however, due to signal and noise interference, peaks of smaller magnitude may also coexist. 
Fig.~\ref{fig:xcorr}(b) exactly reiterates this principle, where the correlation peak is the only useful information, and is representative of the signal's time delay. 
Therefore, our idea is to exploit the underlying \emph{information sparsity in the signal model} to design a simpler acquisition scheme that supports efficient compression, and later recovery. 
In other words, our problem statement is to: \emph{obtain the cross-correlation result (unknown) \textbf{s} using significantly fewer (known) observations of \textbf{x} based on the sparsity sturecture of the problem.}
In the next section, we discuss the theory of sparse approximation and structured sparsity that can exploit this feature.

\subsection{Sparse Approximation and Structured Sparsity} \label{sec:theory}
\noindent
\textbf{Motivation insight.} One can accurately and efficiently recover the information of a high dimensional signal (as $\textbf{x}$) from only a small number of compressed measurements, when the signal-of-interest is sufficiently sparse in a certain transform domain (e.g. \cite{Candes201159}).
\vspace{1mm}
\newline
\noindent
\textbf{The rationale of $\ell^{1}$-minimization.}
Using the sparsifying domain, referred to as a \emph{dictionary} $\Psi \in \mathbb{R}^{n \times d}$ (with full rank), any discrete time signal \textbf{x}~$\in \mathbb{R}^n$ can be represented as a linear combination of columns of $\Psi$ as:
\begin{equation}\label{eq:2}	
		\textbf{x} = \Psi\textbf{s} = \sum^{d}_{i=1} s_{i} \psi_i
\end{equation}
where \textbf{s} $\in \mathbb{R}^d$ is a coefficient vector of \textbf{x} in the $\Psi$ domain, and $\psi_i$ is a column of $\Psi$.
If \textbf{s} is sparse enough, then the solution to an underdetermined system of the form \textbf{x} = $\Psi$\textbf{s} (where the number of unknowns $d$ is greater than the number of observations $n$) can be solved using the following $\ell^{0}$-minimization problem, where the $\ell^{0}$-``norm" counts the number of nonzero entries in a vector.
\begin{equation}\label{eq:3}
	(\ell^{0}): \hspace{0.2cm} \hat{\textbf{s}}_{0} = \arg \min \| \textbf{s} \|_{0} \hspace{0.2cm} \mbox{subject to:} \hspace{0.2cm} \textbf{x} = \Psi\textbf{s} \\
\end{equation}
However, this problem of finding the sparsest solution ($\ell^{0}$-minimization) of an underdetermined system of linear equations is NP-hard \cite{Amaldi1998:l0}.
As an alternative, Candes et al. in \cite{Candes2006:Optimal} and Donoho in \cite{Don:06} show that if $\textbf{s}$ is sparse enough, and $\Psi$ satisfies the Restricted Isometry Property (RIP), then the $\ell^{0}$-minimization problem (Eq.~(\ref{eq:3})) has the same sparse solution as the following  $\ell^{1}$- minimization problem that can be solved in polynomial time by linear programming methods.
\begin{equation}\label{eq:4}
	(\ell^{1}): \hspace{0.2cm} \hat{\textbf{s}}_{1} = \arg \min \| \textbf{s} \|_{1} \hspace{0.2cm} \mbox{subject to:} \hspace{0.2cm} \textbf{x} = \Psi\textbf{s} \\
\end{equation}
However, due to noise (white Gaussian) \textbf{v}~$\in \mathbb{R}^n$ present in real data, \textbf{x} may not be exactly expressed as a sparse superposition of \textbf{s}, and so, Eq.~(\ref{eq:2}) needs to be modified~to:
\begin{equation}\label{eq:5}
		\textbf{x} = \Psi\textbf{s} + \textbf{v} 
\end{equation}
where \textbf{v} is bounded by $\|\textbf{v}\|_{2} < \epsilon$.
The sparse \textbf{s} can still be recovered accurately  by solving the following \textit{stable} $\ell^{1}$- minimization problem via the second-order cone programming.
\begin{equation}\label{eq:6}
	(\ell^{1}_{s}): \hspace{1mm} \hat{\textbf{s}}_{1} = \arg \min \| \textbf{s} \|_{1} \hspace{1mm} \mbox{subject to:} \hspace{1mm} \|\Psi\textbf{s} - \textbf{x}\|_{2} \leq \epsilon   \\
\end{equation}
It is important to note that RIP is only a sufficient but not a necessary condition.
Therefore, $\ell^{1}$-minimization may still be able to recover the sparse \textbf{s} accurately,	even if the sensing matrix $\Psi$ does not satisfy RIP. 
In fact, the use of $\ell^{1}$-minimization to find sparse solutions has a rich history. 
It was first proposed by Logan \cite{Logan:65}, and later developed in \cite{donoho:l1,DonLo:92,CheDoSa:99,DonHu:01,GriNi:03,DonEl:03,ElaBr:02}. 
Here, we use $\ell_1$-minimization to solve the cross-correlation problem via sparse representation.
\vspace{1mm}
\newline
\noindent
\textbf{Dimensionality reduction by random linear projections.} As shown in \cite{BarDaDeWa:08} by the Johnson-Lindenstrauss Lemma, the $\ell^{2}$ distance is preserved in the projection domain with high probability by random projections. 
In other words, all the useful information is preserved in the projection domain. Hence, $\ell^{1}$-minimization can still be used to recover the sparse \textbf{s} from the projected measurements with an overwhelming probability, even though its dimension is significantly reduced. 
More precisely, this projection from high to low dimensional space can be obtained by using a random sensing matrix $\Phi$~$\in \mathbb{R}^{m \times n}$ as:
\begin{equation}\label{eq:7}
		\textbf{y} = \Phi \textbf{x} =  \Phi (\Psi\textbf{s})
\end{equation}
where $m \ll n$ and \textbf{y} $\in \mathbb{R}^m$ is the measurement vector.
In practice, if \textbf{s} has $k \ll d$ nonzero coefficients, then the number of measurements is usually chosen to be \cite{Wright2009:face}:
\begin{equation}\label{eq:8}
		m \geq 2k \log(d/m)
\end{equation}
The sparsity level of \textbf{s} can be verified if the reordered entries of its coefficients decay like the power law; i.e., if \textbf{s} is arranged in the decreasing order of magnitude, then the $d^{th}$ largest entry obeys $|s|_{(d)} \leq Const \cdot d^{-r}$ for $r \geq 1$.
For sparse \textbf{s}, the $\ell^{2}$-norm error between its sparsest and approximated solution also obeys a power law, which means that a more accurate approximation can be obtained with the sparsest \textbf{s}.
However, for efficient recovery, the columns of $\Phi \Psi$ should be as
independent as possible so that the information regarding each
coefficient of \textbf{s} is contributed by a different direction; and this
is achievable if $\Phi$ and $\Psi$ are more incoherent.
Ensembles of random matrices sampled independently and identically (i.i.d.) from Gaussian and $\pm 1$ Bernoulli distributions are largely incoherent with any fixed dictionary, and hence, permit computationally tractable recovery of \textbf{s} \cite{donoho:l1,Candes2006:Optimal}.
\vspace{1mm} 
\newline
\noindent
\textbf{Sparse approximation with structured sparsity.} 
The theory of sparse approximation is applicable to a sensing problem if the underlying signal can be sparsely represented in some dictionary.
A useful feature is that the dimensionality reduction operation is completely independent of its recovery via $\ell^{1}$-minimization.
A sparse signal can be captured efficiently using a limited number of random measurements that is proportional to its information level.
The $\ell^{1}$-minimization process does its best to correctly recover this information with the knowledge of only the dictionary that sparsely describes the signal of interest, when the noise power $\|\textbf{v}\|_{2}$ is small enough and the dictionary $\Psi$ is sufficiently incoherent.
\newline
\indent
The mutual coherence of $\Psi \in \mathbb{R}^{n_{a} \times (2n_{a}-1)}$, denoted as $\mu(\Psi)$, is given as:
\begin{equation}
	\mu(\Psi) = \max_{1 \leq i < j \leq (2n_{a}-1)} \frac{|\Psi_{i}^{T} \Psi_{j}|}{\|\Psi_{i}\|\|\Psi_{j}\|}
\end{equation}
The worst-case coherence $\mu(\Psi)$ corresponds to the largest absolute value of the inner product between two distinct dictionary elements, and is bounded as: $0 \leq \mu(\Psi) \leq 1$.
While it has been proven that any designed $\Psi$ is largely incoherent with $\Phi$, it still may not be \emph{good enough} for parameter estimation - especially under high noise conditions.
Therefore, it is important to prevent coherent pairs of dictionary elements from appearing in the approximation process.
\newline
\indent
Candes et al.\cite[Theorem 1.2]{Candes201159}, in fact, have shown that for conventional sparse approximation with coherent and redundant dictionaries, the reconstruction error is upper bounded by both the noise level and the best $k$-term approximation error.
In another recent work, Duarte and Baraniuk \cite[Theorem 1]{duarte13:spectralcs} have examined that with a structured sparsity model (and using a greedy recovery approach), the upper bound of the reconstruction error decays exponentially to the noise level with an increase in the number of iterations.
Thus motivated by the significant benefits of the structured sparsity model, we propose \emph{StructS-XCorr} that is detailed in the next subsection.

\subsection{Details of StructS-XCorr}

In this section, we present the details of the new dictionary design followed by the computing model of \emph{StructS-XCorr}.

\subsubsection{Design of Representation Dictionary}
\begin{figure}[t]
\begin{center}
\begin{tabular}{c}
\includegraphics[width=3.3in]{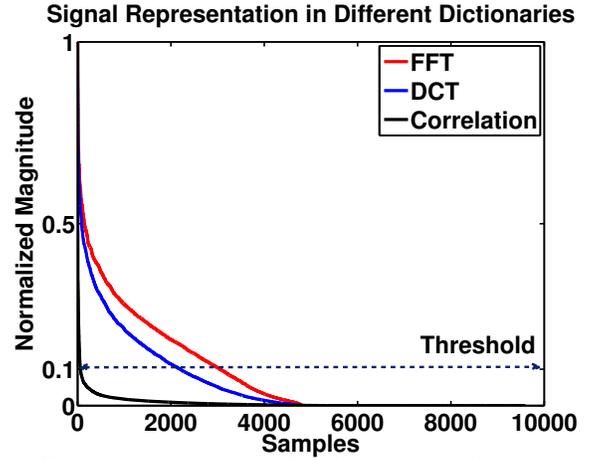}\\
\end{tabular}
\end{center}
\vspace{-8mm}
\caption{\textbf{Signal representation in different dictionaries.} \emph{The signal has a more sparse representation in the correlation dictionary than its FFT and DCT counterparts by an order of magnitude exceeding $2$.}}
\vspace{-2mm}
\label{fig:sparsity}
\end{figure}

\noindent
\textbf{Design guidelines.} The general criteria for designing a reliable
representation dictionary $\Psi$ requires it to sufficiently sparsify the signal $\textbf{x}$.
This one dimension search over the time delay space introduces an important design criteria; where, $\Psi$ should be able to preserve the propagation channel profile information while adhering to the basic design guidelines outlined
by the underlying theory. 
We also define an additional criteria where $\Psi$ should facilitate a faster recovery mechanism that implicitly derives the time delay result without reconstructing the original signal. 
Therefore, the design complexity is to identify and construct a befitting
representation dictionary that satisfies all of the aforesaid requirements.
\vspace{1mm}
\newline
\noindent
\textbf{Design intuition.}
To this end, we were guided by Eq.~\ref{eq:xcorr} where the locally generated reference copy ensembles values from a sweep over all possible (positive and negative) time delay values.
This suggests that the received signal $\textbf{x}$ could be sparsely represented by a single dimension space if we design a representation dictionary having column element that enumerate over all possible time delay combinations.
\vspace{1mm}
\newline
\noindent
\textbf{Design execution.}
For realizing this design goal, we adopt a positive and negative time-shifted
Hankel matrix design of the transmitted signal vector $\textbf{p}$ as $\Psi$.
Note that reversing the time-shifting order results in a Hankel matrix.
We refer to this newly designed $\Psi$ as the \textit{correlation dictionary}.
\newline
\noindent
Depending on the lengths of $\textbf{x}$ and $\textbf{p}$, the following two categories can be identified.
\vspace{0.25mm}
\newline
\noindent
$\bullet$ \emph{Case-1} ($t_{a}$ = $t_{p}$) \textbf{:}
Vectors $\textbf{p}$ and $\textbf{x}$ are of equal dimensions with $n_{a}$ samples.
The elements of $\Psi \in \mathbb{R}^{n_{a} \times (2n_{a}-1)}$ are given as:
\begin{equation}\label{eq:12}
\Psi(:,i) = 
\begin{cases}
\left[ zeros(n_{a}-i) \hspace{1mm} p(1:i) \right]^T & \hspace{-2mm} 1 \leq i \leq n_{a} \\
\left[ p(i+1-n_{a}:n_{a}) \hspace{1mm} zeros(i-n_{a}) \right]^T & \\
& \hspace{-2.5cm} (n_{a}+1) \leq i \leq (2n_{a}-1) \\
\end{cases}
\end{equation}
where $\Psi(:,i)$ denotes the $i$th column, $\left[ \cdot \right]$ denotes a vector of length $n_{a}$, 
$zeros(i)$ denotes a zero vector of length $i$, $\cdot^T$ denotes the transpose of a vector (matrix), and $p(i:j)$ denotes a vector of elements with indices from $i$ to $j$ of the input sample set $\textbf{p}$.
\newline
\noindent
$\bullet$ \emph{Case-2} ($t_{a} > t_{p}$) \textbf{:}
The size of $\textbf{x}$ is greater than $\textbf{p}$, and so, the system is balanced by right zero-padding $(n_{a}-n_{p})$ entries~to~$\textbf{p}$.
\newline
\indent
Other popular dictionaries such as the FFT and DCT, in contrast, do not provide as good a sparse depiction as the proposed correlation dictionary, and also, do not satisfy the remaining two requirements (important for ranging).
Fig.~\ref{fig:sparsity} compares their sparsity levels (for an indoor high multipath channel) by sorting the samples by their magnitudes.
The fastest decay characteristic (or the smallest $k$) is observed in the correlation domain, and so, offers the most sparse representation.
This implies that the most accurate approximations (or range estimates) can be obtained with the correlation dictionary using the smallest number of measurements $m$ (Eq.~(\ref{eq:8})).

\begin{figure*}[t]
\begin{center}
\subfigure[Standard cross-correlation]{\label{fig:reconstruction_without_buffers_a}\includegraphics[width=2.2in]{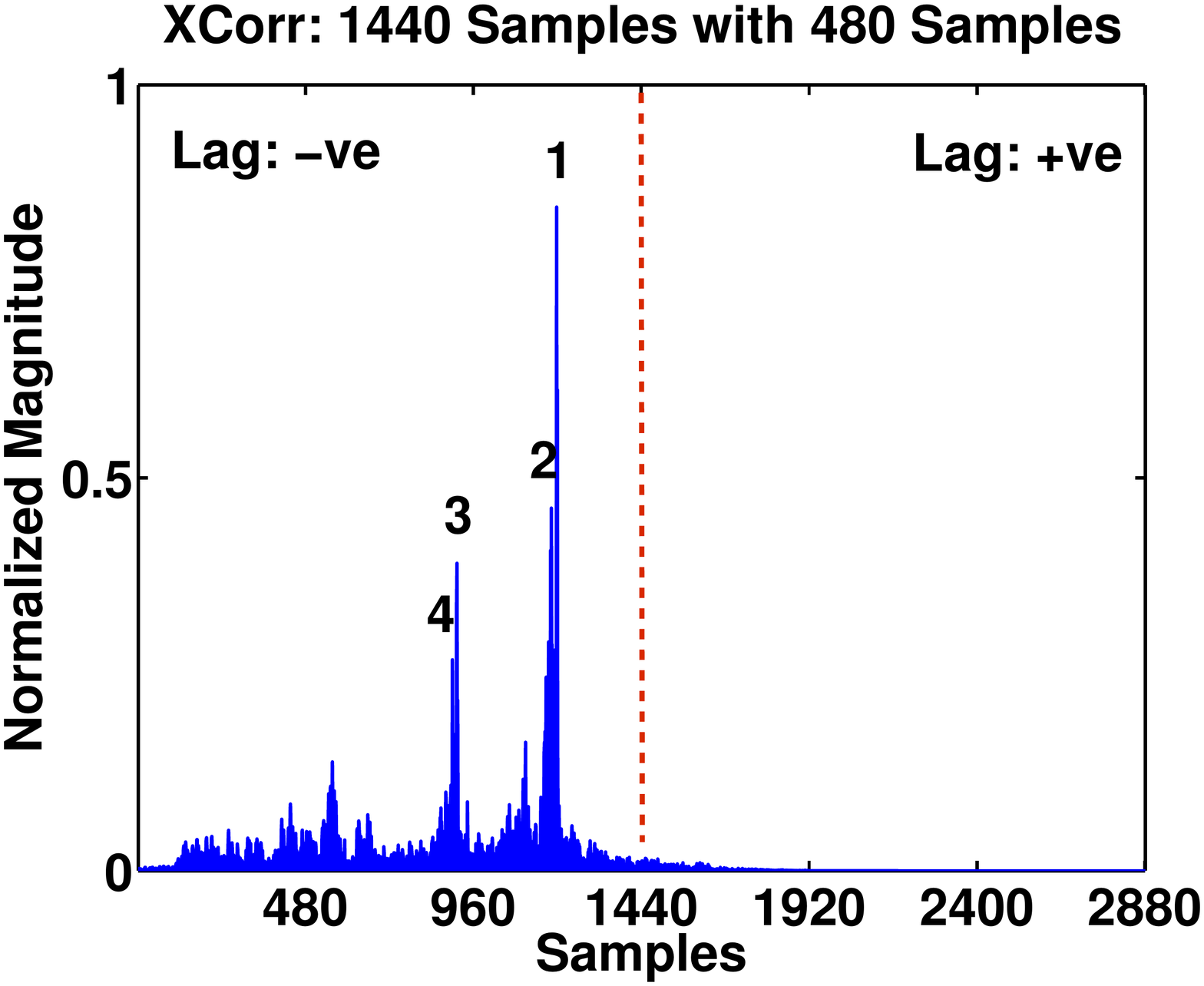}}
\subfigure[Recovery via sparse approx.]{\label{fig:reconstruction_without_buffers_b}\includegraphics[width=2.2in]{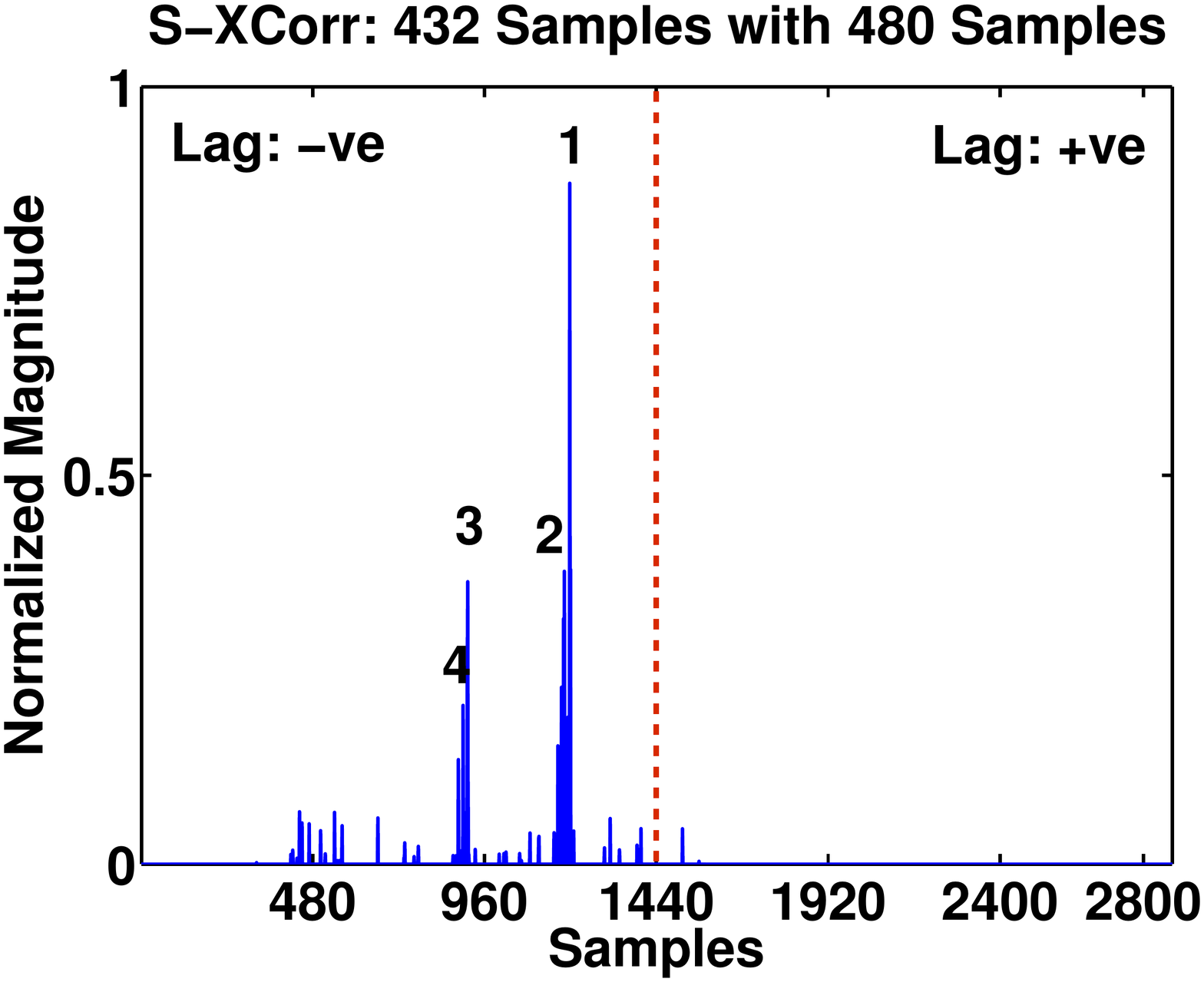}}
\subfigure[Recovery via structured sparse approx.]{\label{fig:reconstruction_without_buffers_c}\includegraphics[width=2.2in]{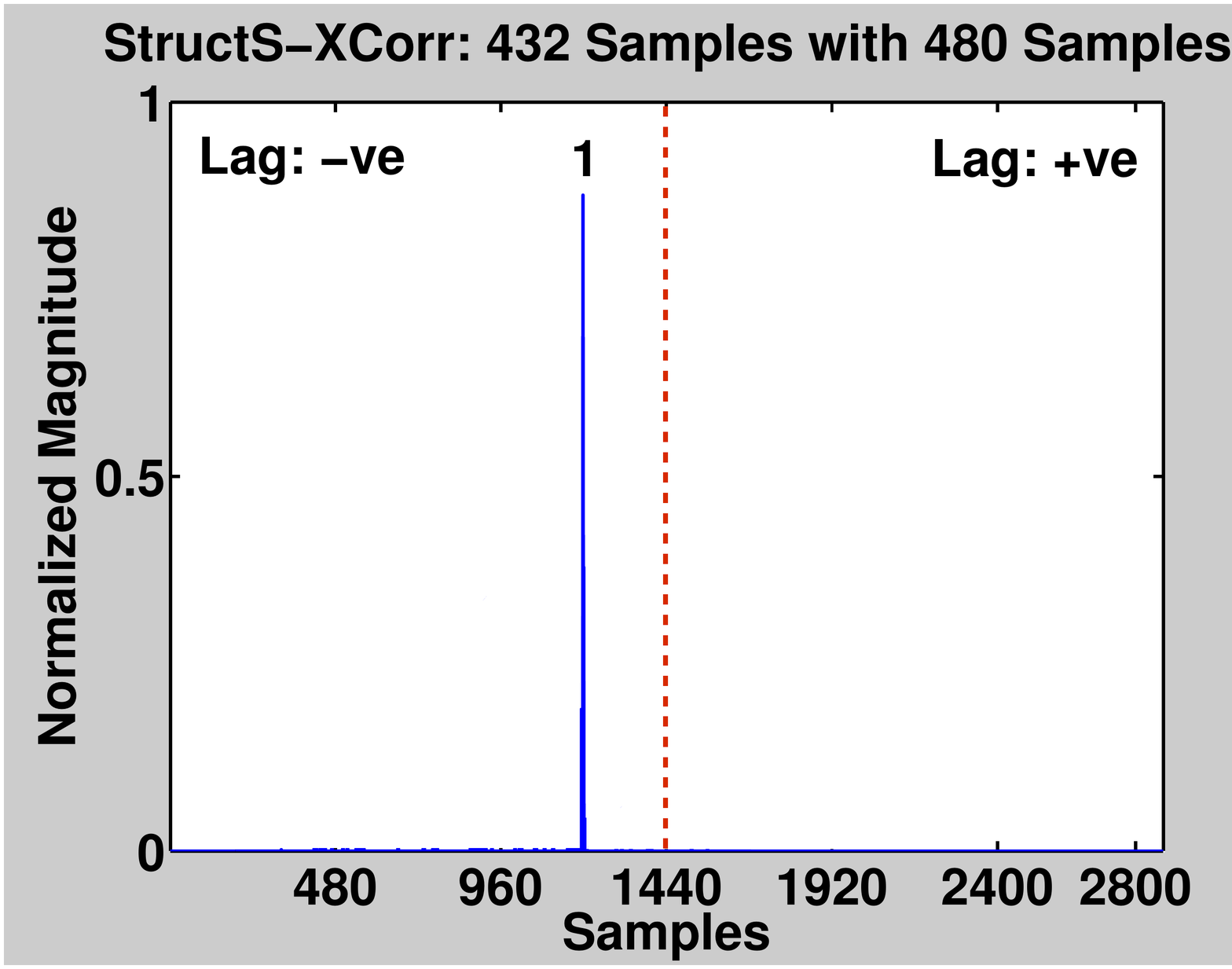}}
\end{center}
\vspace{-2mm}
\caption{\textbf{Validation: \emph{StructS-XCorr} vs. \emph{\{S-XCorr, XCorr\}}.} \emph{The detection accuracy of StructS-XCorr is on a par with S-XCorr and XCorr, but with better robustness against multipaths and low-noise peaks.}} 
\label{fig:reconstruction_without_buffers}
\end{figure*}

\subsubsection{Compression and Recovery} \label{sec:comp_n_recovery}
\noindent
\textbf{Compression.}
The dimensions of \textbf{x} $\in \mathbb{R}^{n_{a}}$ are significantly reduced at the receiver by multiplying it with a random sensing matrix $\Phi$ $\in \mathbb{R}^{m \times n_{a}}$ resulting in the measurement vector \textbf{y} $\in \mathbb{R}^{m}$ ($m \ll n_{a}$) as: $\textbf{y} = \Phi \textbf{x}$.
$m$ is related to $n_{a}$ by the compression factor $\alpha$ given as: $m = \alpha \hspace{1mm} n_{a}$ where $\alpha \in \left[ 0,1\right]$.
For example, $\alpha=0.10$ means that the information in \textbf{x} has been compressed by $90\%$. $\Phi$ is a binary sensing matrix with its entries identically and independently (i.i.d.) sampled from a balanced symmetric Bernoulli distribution of $\pm 1$.
\begin{equation}\label{eq:11}
\Phi =  \frac{1}{\sqrt{m}} \bar{\Phi} \hspace{3mm} \mbox{where} \hspace{1mm} \bar{\Phi}_{i} \hspace{1mm} \mbox{i.i.d.} \hspace{2mm} \textbf{Pr}(\bar{\Phi}_{i,j} = \pm 1) = 0.5
\end{equation}
Binary ensembles have a shorter memory representation than Gaussian ensembles, and also, alleviate operational complexity; hence, they are economical for sensor platforms.
A balanced $\Phi$ consists of $\pm 1$ at equal probability, where each row contains equal number of $1$'s and -$1$'s. 
Therefore, in each row of $\Phi$, the sum of the elements is always zero. 
A balanced $\Phi$ provides a higher probability of detection (at recovery) if the noise in x is Gaussian{\cite{Shen2012:EBS}.
The receiver transfers $m$ samples of \textbf{y} to the base-station (BS) for post-processing.
\begin{figure}[b]
\begin{center}
\begin{tabular}{c}
\includegraphics[width=3.3in]{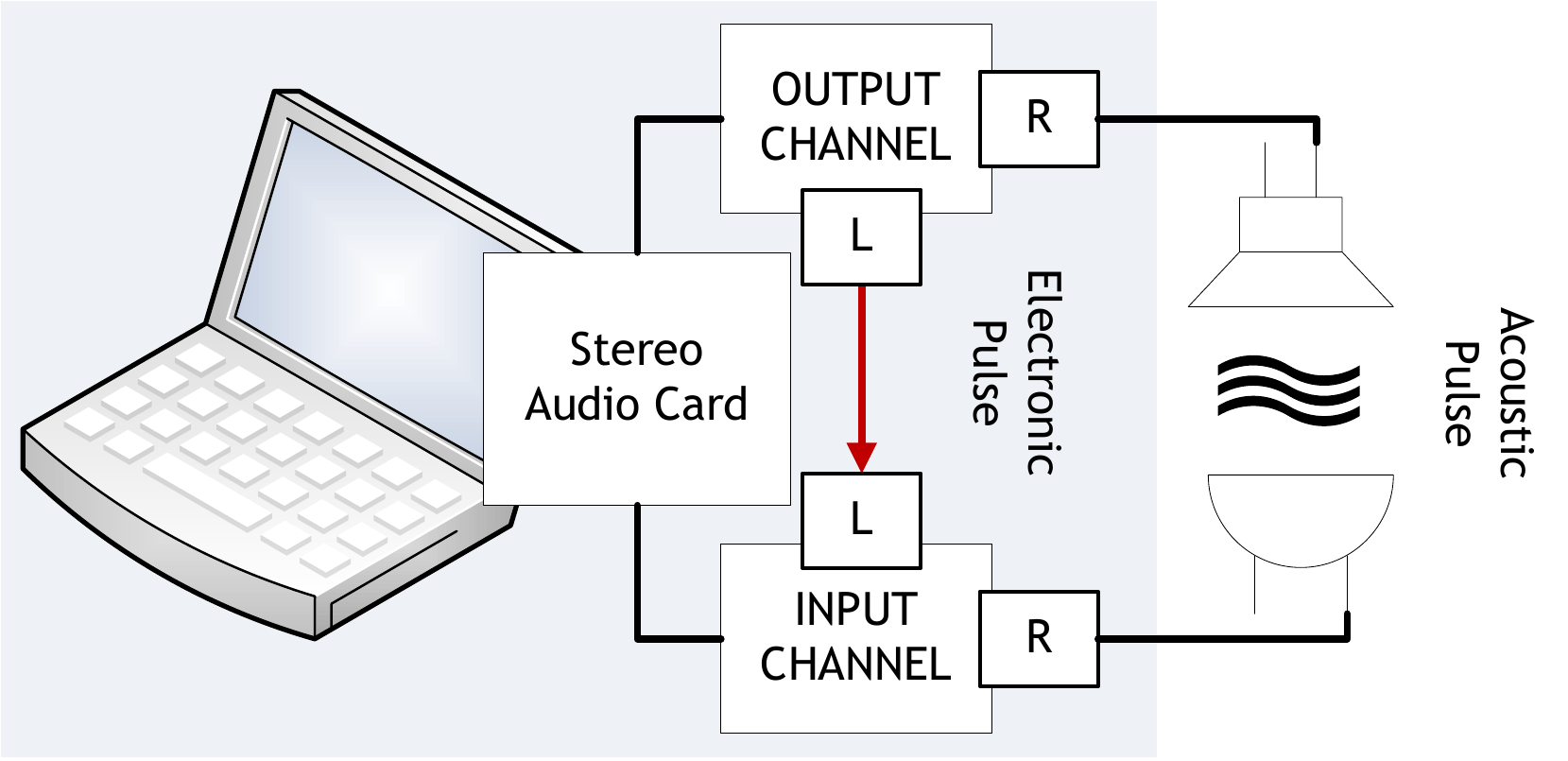} \\
\end{tabular}
\end{center}
\vspace{-2mm}
\caption{\emph{System architecture of a custom designed acoustic ranging system for empirical characterization of StructS-XCorr.}}
\label{fig:poc_system}
\end{figure}
\vspace{1mm}
\newline
\noindent
\textbf{Recovery via \emph{Sparse Approximation (S-XCorr)}\cite{Misra2012:sparsexcorr}.}
The BS uploads the compressed measurements to a service application on the control server. 
It requires the a-priori knowledge of the seed that generates $\Phi$, and the dictionary $\Psi$.
Since $\textbf{x}$ can be represented sparsely as $\textbf{s}$ in the dictionary $\Psi$  and  \textbf{x} (the received signal) is known, the desired sparse solution \textbf{s} can be recovered by solving Eq.~(\ref{eq:6}).
However, as the dimensions of \textbf{x} are reduced significantly via Eq.~(\ref{eq:7})\footnote{\scriptsize Direct cross-correlation in the projection domain (using \textbf{y}) did not produce desirable ranging results because \textbf{y} consists of random projections.}, the \textit{reduced} $\ell^{1}$- minimization problem for a given tolerance $\epsilon$ is given as:
\begin{equation} \label{eq:12}
	(\ell^{1}_{r}): \hspace{1mm} \hat{\textbf{s}}_{1} = \arg \min \| \textbf{s} \|_{\ell_{1}} \hspace{1mm} \mbox{s.t:} \hspace{1mm} ||\Phi \Psi \textbf{s} - \textbf{y}||_{2} \leq \epsilon
\end{equation}
\newline
\noindent
$(\ell^{1}_{r})$ is known as Lasso\footnote{\scriptsize The minimizer of $\| x - \Psi s \|^{2}_{2} + \lambda \| x \|_{1}$ is defined as the Lasso solution; where $\lambda$ can be referred as the inverse of the Lagrange multiplier associated with a constraint $\| x - \Psi s \|^{2}_{2} \leq \epsilon$. 
For every $\lambda$, there is an $\epsilon$ such that the two problems have the same solution.}
in statistical literature, and regularizes highly undetermined linear systems when the desired solution is sparse.
\emph{The correlation domain coefficients $\hat{\textbf{s}}_{1}$ are related to the various propagation (direct and reflected) paths, where the index of the first tallest correlation coefficient peak is the estimate of the pulse arrival time of the direct path, and thus, provides the range.}
\vspace{1mm}
\newline
\noindent
\textbf{Recovery via \emph{Structured Sparsity and Sparse Approximation (StructS-XCorr)}.}
In our case, $\Psi$ is not strictly incoherent due to the repetitive
nature of the elements along each row of the matrix (which is an artifact of Hankel matrices).
\begin{algorithm}
\caption{Structured SparseXcorr $\textbf{s}_s =
  \mathcal{S}_k(\textbf{s}, \mu_0)$}
\label{algo:structuredsparsexcorr}
\begin{algorithmic}[1]
\REQUIRE Coefficient vector \textbf{s}, target coherence $\mu_{0}$
\ENSURE Structured sparse coefficient vector $\textbf{s}_{s}$
\STATE Initialize $\textbf{s}_{s}=\textbf{0}$, $i=1$
\WHILE{$i < k$ and $\textbf{s} \neq \textbf{0}$}
			\STATE $l^{\star} = \arg\max_{1 \leq l \leq 2n-1}|s(l)|$
			\STATE $\textbf{s}_{s}(l^{\star})=\textbf{s}(l^{\star})$
			\STATE $\Lambda = \{\lambda: \frac{|\psi_\lambda^T
                          \psi_{l^*}|}{\|\psi_\lambda\| \|\psi_{l^*}\|} \geq
                        \mu_{0} \}$ 
			\STATE $\textbf{s}|\Lambda = \textbf{0}$
			\STATE $i = i + 1$
\ENDWHILE
\end{algorithmic}
\end{algorithm}
\noindent
To overcome the shortcomings due to dictionary coherence, we apply the principles of structured sparsity to the \emph{S-Xcorr} computing model.
Although the optimal solution can be obtained via linear programming, we adopt a computationally efficient heuristic\cite{duarte13:spectralcs} after executing Eq.~\ref{eq:12}.
It is presented in Procedure~\ref{algo:structuredsparsexcorr} that works as follows. 
In line~$3$ and line~$4$ of the procedure, we select the entry $s(l^*)$ of the coefficient vector $\textbf{s}$ with largest magnitude that is then pushed into the output of the structured sparse vector \textbf{s}.
To prevent coherent elements from appearing simultaneously in $\Psi s_s$, we define (line~$5$) the set $\Lambda$ of all indices $\lambda$ for which the inner product of $\psi_\lambda$ and $\psi_{l^*}$ is larger than some predefined threshold $\mu_0$.
This set indices all dictionary elements that are coherent with the newly selected one, and their future selection is prevented in line~$6$ by setting the corresponding entries of the vector $s$ to zero.

\subsection{Analysis of StructS-XCorr} \label{sec:analysis-structs-xcorr}
In this section, we analyze the performance of \emph{StructS-Xcorr} and identify challenges in detection reliability.
\vspace{1mm}
\newline
\noindent
\textbf{Experimental system.}
We conducted this study using a custom designed acoustic ranging system (Fig.~\ref{fig:poc_system}) with different assembled units.
The front-end of the transmitter consisted of a COTS ribbon (speaker) transducer, but driven by an custom assembled (external) wideband power amplifier with a tunable ($5$-$20$ times) gain controller.
The receiver front-end comprised of a custom designed preamplifier unit interfaced with a COTS Knowles microphone (SPM$0404$UD$5$).
The controller for this system was setup on a laptop, where synchronization and ranging signals were generated, captured and analyzed for range estimation.
The operational sequence commenced with the generation of the linear chirp [$01$-$20$]\,kHz/$0.01$\,s that was then directed into two separate streams: \emph{first}, left input channel of the ADC of the audio card (i.e., an electronic chirp) and \emph{second}, wideband amplifier (i.e., an acoustic chirp).
The electronic chirp is equivalent to an RF pulse and marks the transmission time of the acoustic chirp, which is thereafter detected by the receiver unit and directed into the right input channel of the ADC.
The received acoustic signal is considered from the time marker provided by the electronic chirp so as to discard the delays incurred during the transmission stage\footnote{The experimental setup mimics the concept of velocity-difference TDOA (V-TDOA)\cite{misra2015:acoustic_rangefinders}}. 
At the processing station (that functionally replicates the receiver post-processing stage and BS), the acquired samples are first compressed and subsequently recovered to estimate the range.

\subsubsection{Ranging Challenges and Mitigation} \label{sec:detection}
\vspace{1mm}
\noindent
\textbf{Analysis: basic ranging performance.}
In this experimental setup, the transmitter and the receiver were placed $1.5$\,m apart.
The ranging process was performed with the receiver configured to record for $0.03$\,s - just long enough to capture the ranging signal along with its multipaths.
The audio card was configured to sample at $48$\,kHz; hence, the transmitted signal \textbf{p} and the acquired trace \textbf{x} consisted of $480$ and  $1440$ samples respectively.
Using $\alpha$ = $0.30$, \textbf{x} was compressed to obtain the measurement vector \textbf{y} of $432$ samples followed by its recovery to obtain \textbf{s} (Section~\ref{sec:comp_n_recovery}) using \emph{S-XCorr} and \emph{StructS-XCorr}, and its accuracy is then validated against \emph{XCorr} (Eq.~\ref{eq:maxlikelihood}).
Fig.~\ref{fig:reconstruction_without_buffers}(a), Fig.~\ref{fig:reconstruction_without_buffers}(b) and Fig.~\ref{fig:reconstruction_without_buffers}(c) show the respective results, where we observe that \emph{all} the methods obtain exactly the same estimate for the position of the first tallest peak at a negative lag of $220$ samples.
\emph{S-XCorr} is able to obtain the multipath profile\footnote{The generation of the dictionary coefficients and cross-correlation peaks are in the negative lag part since we have reversed the order of operation, wherein the reference signal was operated with the acquired signal.}, but it is not accurate as it does not follow the same height-to-position relationship (observe the position of peak-$2$ \& $3$ in Fig.~\ref{fig:reconstruction_without_buffers}(b) as suggested by the corresponding \emph{XCorr} result shown in Fig.~\ref{fig:reconstruction_without_buffers}(a)).
Although, these parameters are not important for distance estimation, they are - nevertheless - legitimate sources of erroneous detection.
\emph{StructS-XCorr}, on the other hand, does not recover multipath/low-noise peaks apart from the LoS path (Fig.~\ref{fig:reconstruction_without_buffers_c}); and therefore, alleviates such anomalies.
\vspace{1mm}
\newline
\noindent
\textbf{Analysis: space and time complexity.}
The functionality algorithm (\emph{XCorr vs. compression}) on the receiver is the vital point of difference.
The running time of \emph{XCorr} is O$(n^{2})$ in the time domain (\emph{TD-XCorr}) and O$(n \log n)$ in the frequency domain (\emph{FD-XCorr}) on conventional receiver systems.
However, for WSN nodes, additional signal processing platforms have to be added to compensate for the lack of hardware divide or floating point support units.
Therefore, we propose an alternate data compression functionality that has a similar time complexity ($mn$ $\approx$ O$(n \log n)$), but a much smaller space complexity (competent with the mote constraints).
\newline
\indent
We compared their performance on the experimental system, for which we performed the same ranging process but configured the receiver to record for $0.1$\,s (i.e., 4800 acquired samples).
Table \ref{tab:pertestPC} shows the individual running time of the \emph{TD-XCorr}, \emph{FD-XCorr} and compression for different compression factors $\alpha$.
We note that \emph{FD-XCorr} is $\approx$ $30$ times faster than \emph{TD-XCorr} as expected from their asymptotic results.
However, the compression time (shown as `Compression 1-Buf') varies for different $\alpha$, and is slower than \emph{FD-XCorr} for all except $\alpha=0.05$.
\newline
\indent
We overcome this drawback by using the simple idea of \emph{buffer-by-buffer} compression rather than one-step compression.
This method divides the acquired signal vector $\textbf{x}$ of length $n_{a}$ across $b$ buffers of equal sizes, compresses the information in each buffer, and finally, assembles the measurements in their correct order.
The signal in each buffer $\tilde{\textbf{x}}$ is of length $\tilde{n}$, where $\tilde{n} = n_{a} / b $.
The random sensing matrix $\Phi$ for compressing the data in each buffer is of size $\left[ \tilde{m} \times \tilde{n} \right]$, where $\tilde{m} = \alpha \hspace{1mm} \tilde{n} = \alpha \hspace{1mm} (n_{a} / b) = \hspace{1mm} m / b$.
The resultant measurement vector $\tilde{\textbf{y}}$ (for each buffer) is of length of $\tilde{m}$.
The number of iterations required to process each buffer is ($\tilde{m}\tilde{n}$).
Therefore, the compression time for $b$ buffers take ($b\tilde{m}\tilde{n}$) = $(mn_{a}/b)$ iterations.
This improvement can be identified in Table \ref{tab:pertestPC} (shown as `Compression 10-Buf'), where we divide the $4800$ samples across $10$ buffers and record their individual compression time for different $\alpha$.
The results show a worst-case to best-case improvement of $6 \times$ to $60 \times$ over \emph{FD-XCorr}. 
As resource constrained WSN motes do not support floating point operation, our proposed method is expected to yield better performance (shown in Section~\ref{sec:evaluation}) on such platforms than on a standard PC.
\begin{table}[t]
\begin{center}
\caption{Time Complexity Analysis}
\begin{footnotesize}
\begin{tabular}{p{0.5cm} p{1.3cm} p{1.3cm} p{1.5cm} p{1.75cm}}
\toprule
	\textbf{$\alpha$} & \textbf{TD-XCorr} & \textbf{FD-XCorr} &  \textbf{Comp.} & \textbf{Comp.}\\ 
	\textbf{} & \textbf{(s)} & \textbf{(s)} &  \textbf{1-Buffer (s)} & \textbf{10-Buffers (s)}\\ 
\midrule
	\textbf{0.05}	  & \textbf{0.1932} &	\textbf{0.0062} &	\textbf{0.0042} &	\textbf{0.0001} \\
  0.10	  & 0.1932 &	0.0062 &  0.0077 &	0.0003 \\  
  \textbf{0.30}	  & \textbf{0.1932} &	\textbf{0.0062} &	\textbf{0.0218} &	\textbf{0.0006}\\   
  0.50	  & 0.1932 &	0.0062 &	0.0361 &	0.0010 \\
\bottomrule
\label{tab:pertestPC}
\end{tabular}
\end{footnotesize}
\end{center}
\vspace{-3mm}
\end{table}
\vspace{1mm}
\newline
\noindent
\textbf{Analysis: signal detection and post-processing.}
The process of detection is not without errors as the reconstructed coefficients $\textbf{s}$ may have been wrongly approximated due to measurement noise that contributes to higher coefficient values at incorrect locations.
To overcome these inaccuracies, we use the same principle of buffer-by-buffer reconstruction at the BS as well, which not only provides an additional clue for correct detection, but also, serves as a guideline to choose the buffer count $b$.
\newline
\indent
The number of buffers $b$ is chosen such that the number of samples in each buffer is the same as the sample count of the reference signal $\textbf{p}$, i.e., $\tilde{n} = t_{p}F_{s}$.
For example, if $\textbf{p}$ contains $100$ samples and $\textbf{x}$ consists of $1000$ samples, then $b$ is $10$.
There are two benefits in making this choice.
First, it restricts the direct path signal (in the total acquired trace) to be spread across a maximum of $2$ buffers, and so, guarantees that the magnitude of the corresponding recovered coefficient would always remain at least $50\%$ above its original estimate. 
Increasing $b$ beyond $2$ buffers decreases the individual peak heights to smaller magnitudes that poses a difficult detection task to differentiate them from the noise-floor.
Second, it provides easy processing at the BS, where the operation of right zero-padding $\textbf{p}$ to make its dimensions equal to $\textbf{x}$ is substituted by fragmenting $\textbf{x}$ into $b$ buffers to match the size of $\textbf{p}$ (Section~\ref{sec:comp_n_recovery}).
\newline
\indent
The reconstruction process is performed on all $b$ buffers, which is followed by the signal detection and range estimation algorithm.
\vspace{1mm}
\newline
\noindent
$\bullet$ \emph{Phase-1:}
It identifies the various correlation domain coefficient peaks and selects the first tallest peak in each of the $b$ buffers that is at least $6$ standard deviations above the mean.
The detection is considered to have failed for those buffers where no point qualifies as a peak.
This reduces the validation space for phase-$2$ to $\tilde{b}$ ($\leq b$) buffers.
\vspace{1mm}
\newline
\noindent
$\bullet$ \emph{Phase-2:} 
If there are valid peaks in more than one buffer (i.e., $\tilde{b} > 1$), then the tallest peak (across all $\tilde{b}$ buffers) among them is selected as the ranging peak. 
The detection is correct, if this peak in buffer $b_{i}$ has a lag that is:
\begin{itemize}
	\item Positive: $\Rightarrow$ The peak in the previous buffer $b_{i-1}$ must have a negative lag.
	\item Negative: $\Rightarrow$ The peak in the next buffer $b_{i+1}$ must have a positive lag.
\end{itemize}
\noindent
This relationship is a result of the manner in which the signal gets aligned in different buffers and its equivalent representation in the correlation domain/cross-correlation (Fig.~\ref{fig:reconstruction_with_buffers}).
\newline
If $\tilde{b} = 1$ (i.e., only a single buffer has a valid peak), then the peak identified in phase-$1$ qualifies as the ranging peak.
The estimated range $r$ is obtained as:
\begin{equation}
	r = ((\tilde{n} b_{i-1} + \hat{l})/F_{s}) \times v_{s}
\end{equation}
where $b_{i-1}$ is the buffer count before the detection buffer, $\hat{l}$ is the lag (in samples) of the ranging peak in the detection buffer, and $v_{s}$ is the temperature compensated speed of sound in air.

\begin{figure}[t]
\begin{center}
\subfigure[]{\label{fig:reconstruction_with_buffers_c}\includegraphics[width=1.65in]{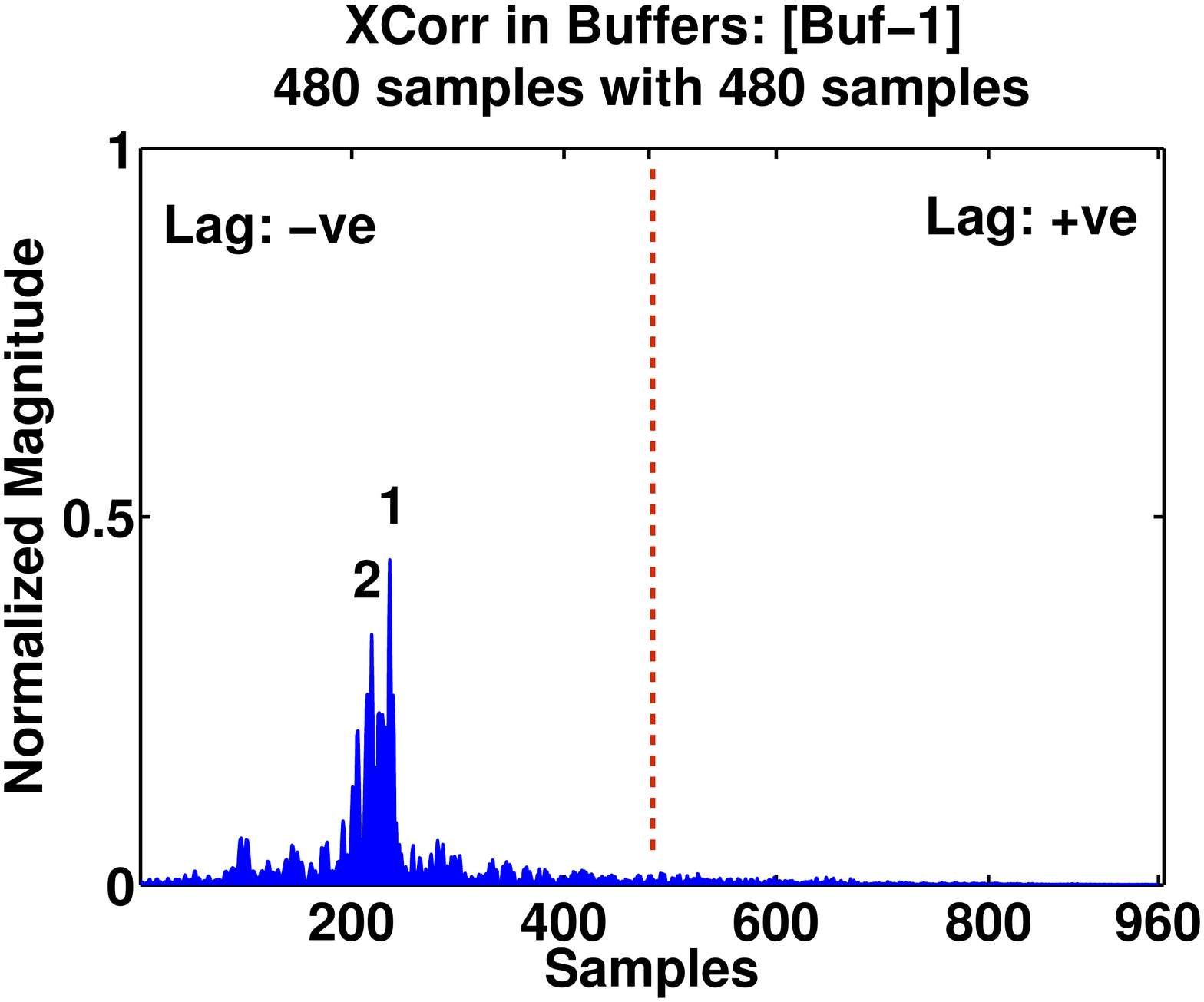}}
\subfigure[]{\label{fig:reconstruction_with_buffers_d}\includegraphics[width=1.65in]{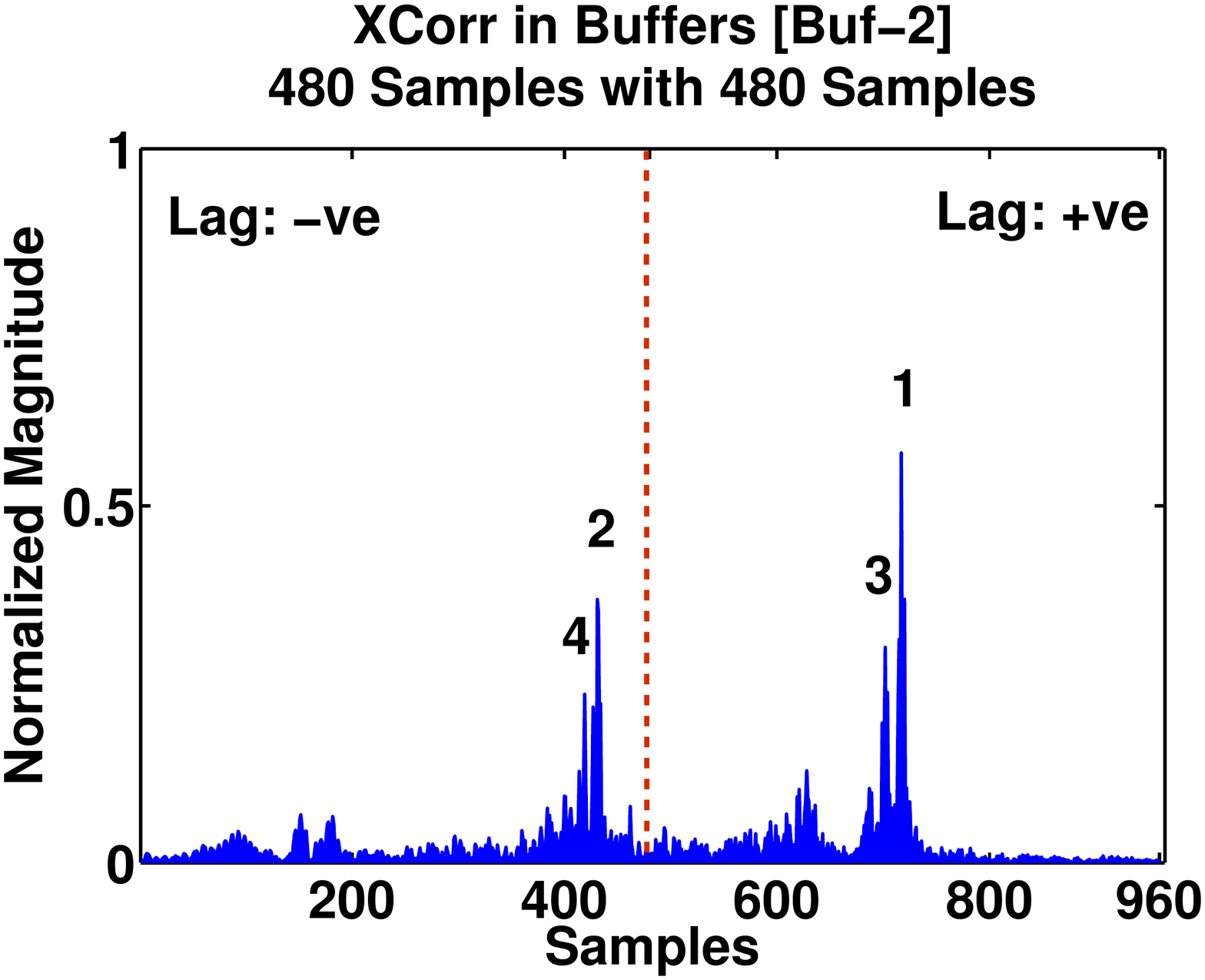}}
\subfigure[]{\label{fig:reconstruction_with_buffers_a}\includegraphics[width=1.65in]{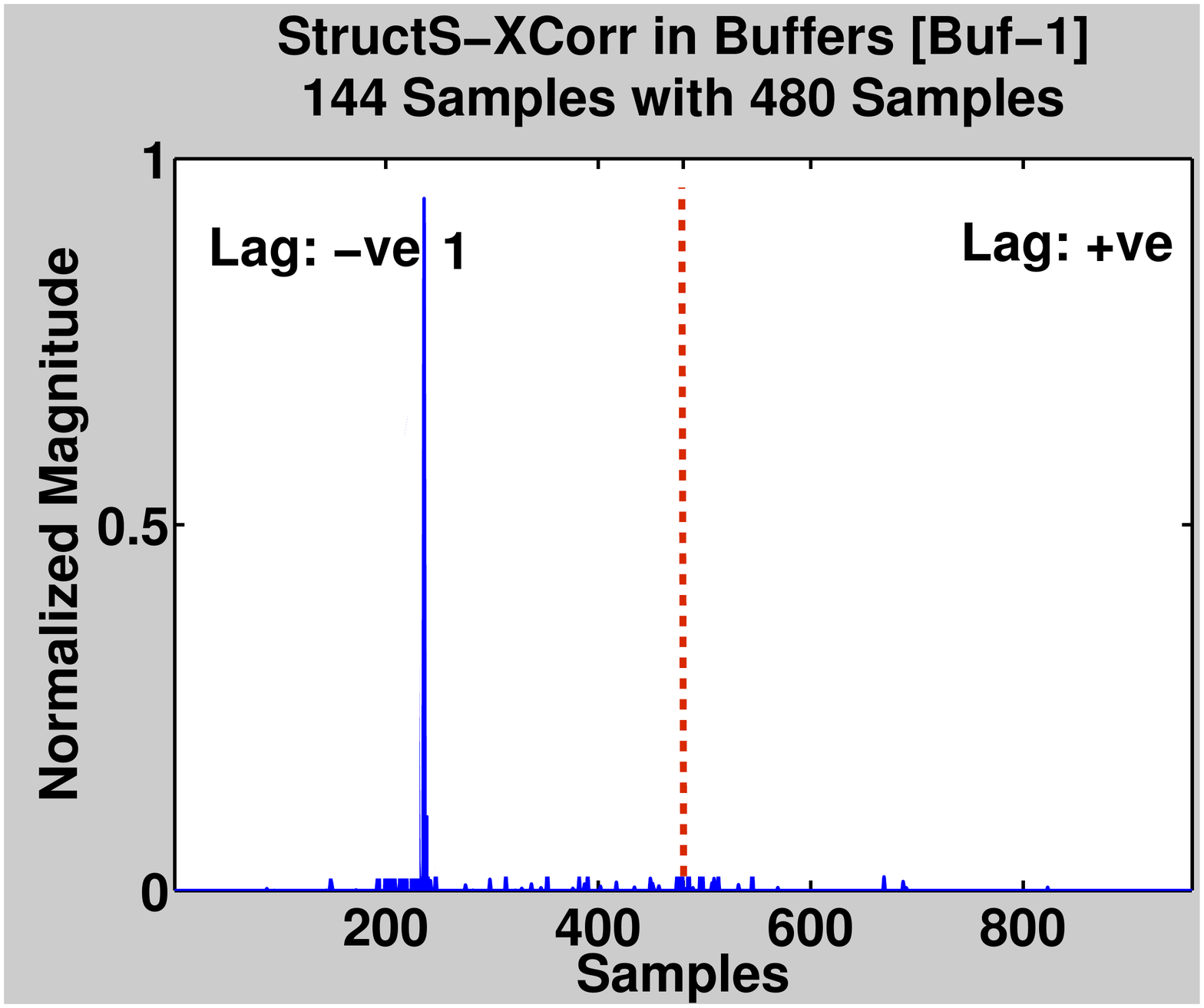}}
\subfigure[]{\label{fig:reconstruction_with_buffers_b}\includegraphics[width=1.65in]{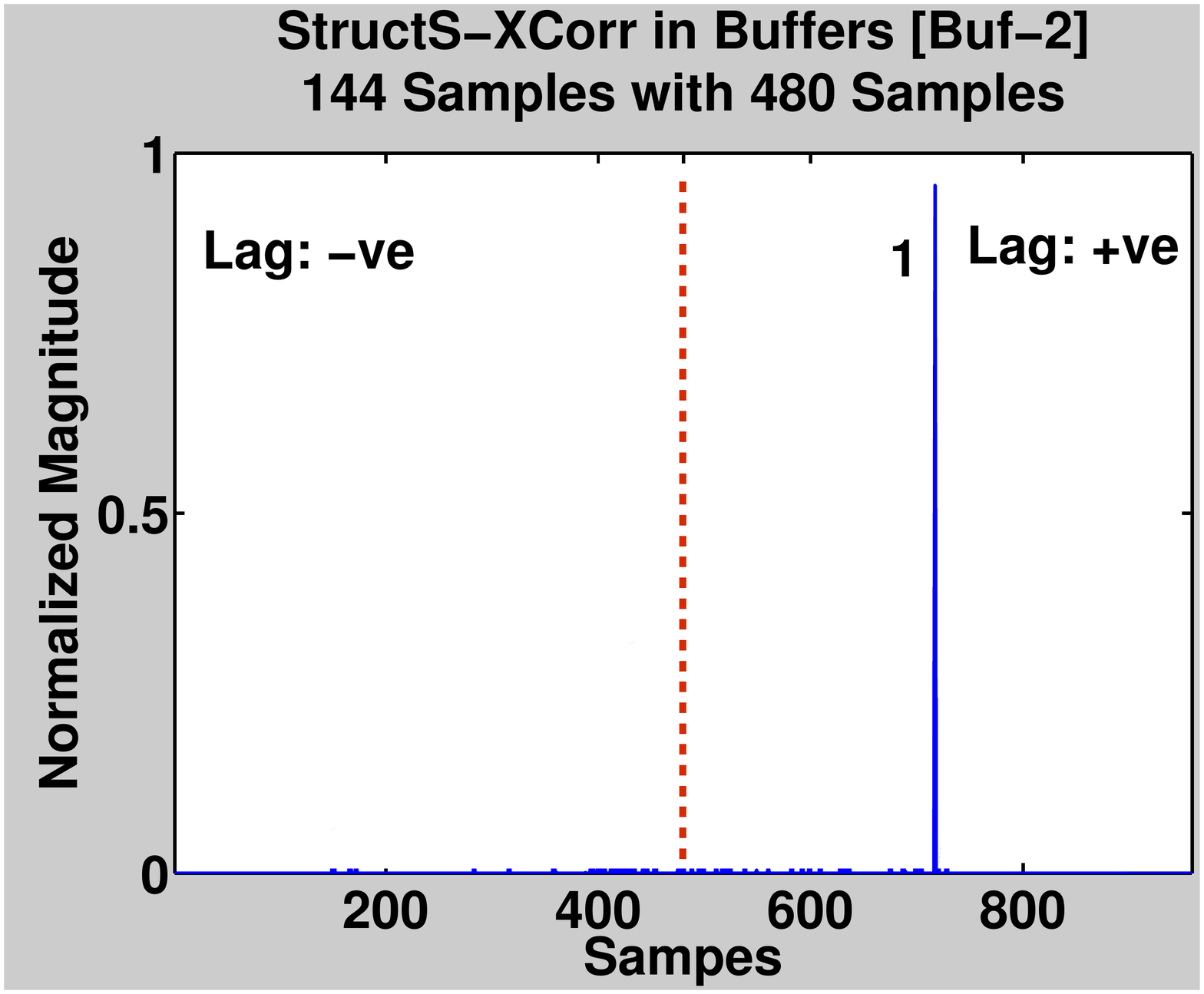}}
\end{center}
\vspace{-4mm}
\caption{\textbf{Buffer-by-Buffer Processing.} \emph{The detection accuracy of StructS-XCorr, in regards to the position of the LoS peak and the tallest peak in each buffer, is at par with XCorr.}}
\vspace{-2mm}
\label{fig:reconstruction_with_buffers}
\end{figure}
\begin{figure*}[t]
\begin{center}
\begin{tabular}{ccc}
\includegraphics[width=2.25in]{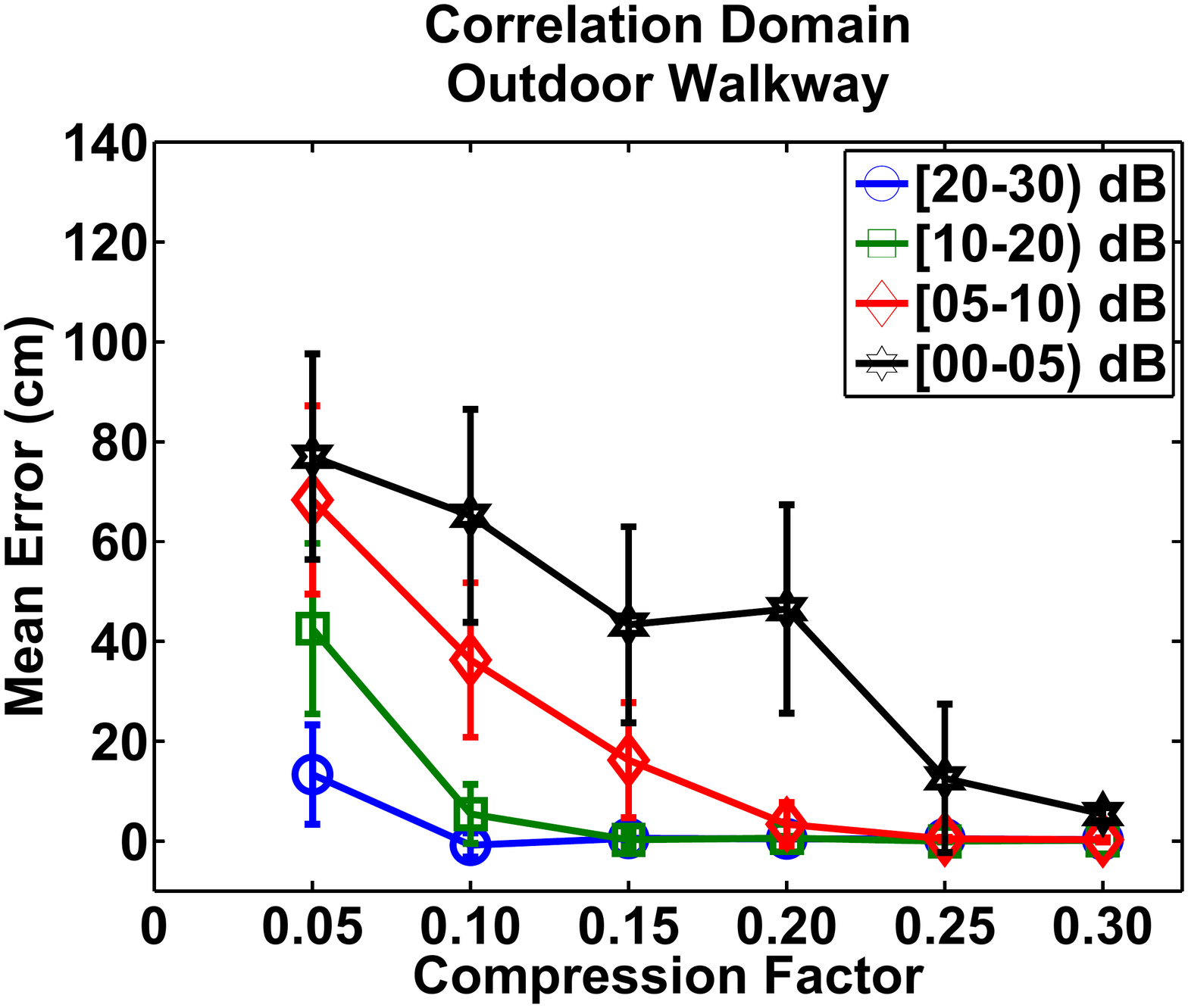} & \includegraphics[width=2.25in]{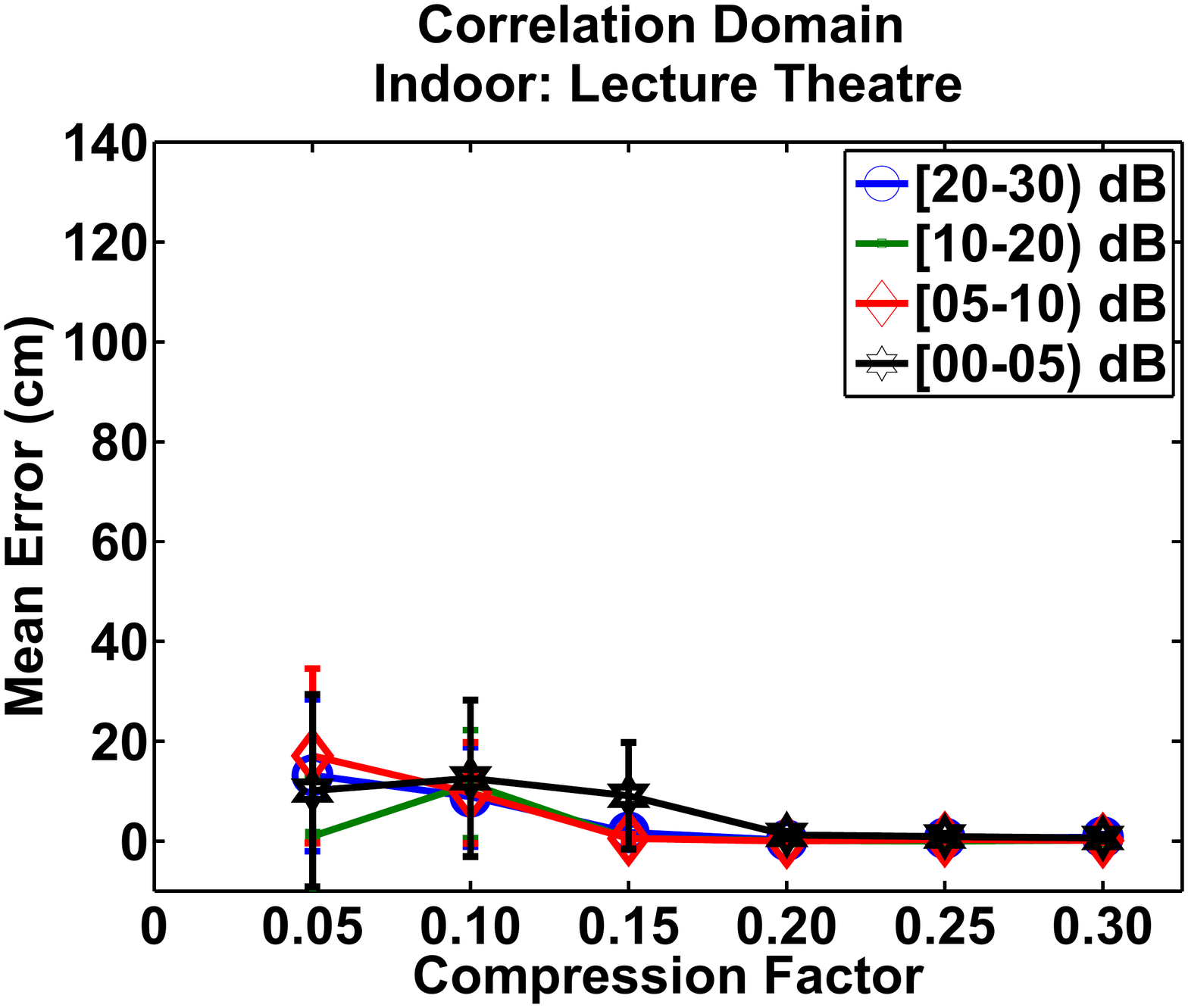} & \includegraphics[width=2.25in]{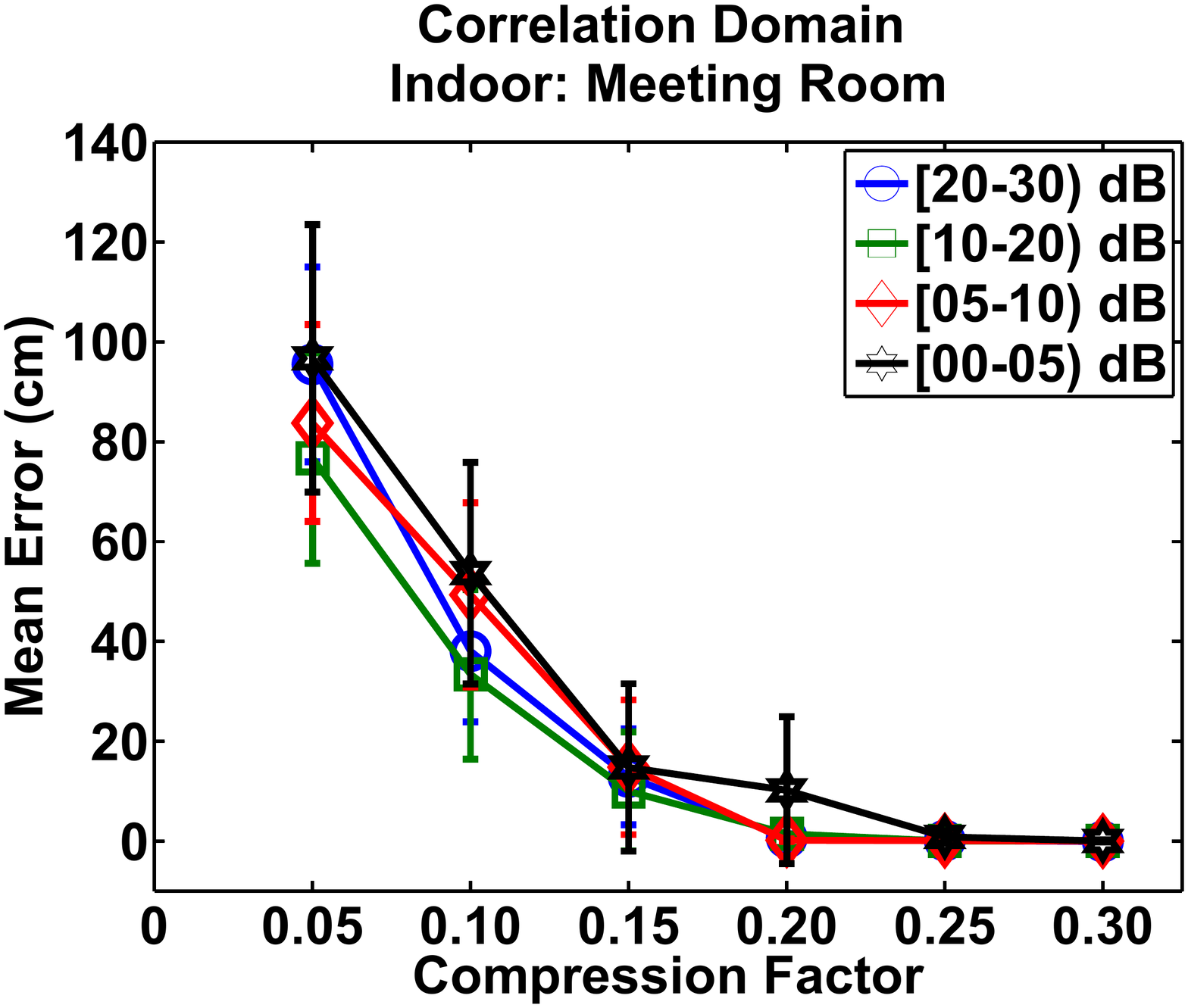}\\
(a) Case-A:  & (b) Case-B: & (c) Case-C: \\
Multipath: very-low & Multipath: low & Multipath: high\\
\end{tabular}
\end{center}
\vspace{-3mm}
\caption{\textbf{S-XCorr.} \emph{Characterization of compression factor $\alpha$ with SNR.}}
\label{fig:fixed_dis_variable_power}
\end{figure*}
\begin{figure*}[t]
\begin{center}
\begin{tabular}{ccc}
\includegraphics[width=2.25in]{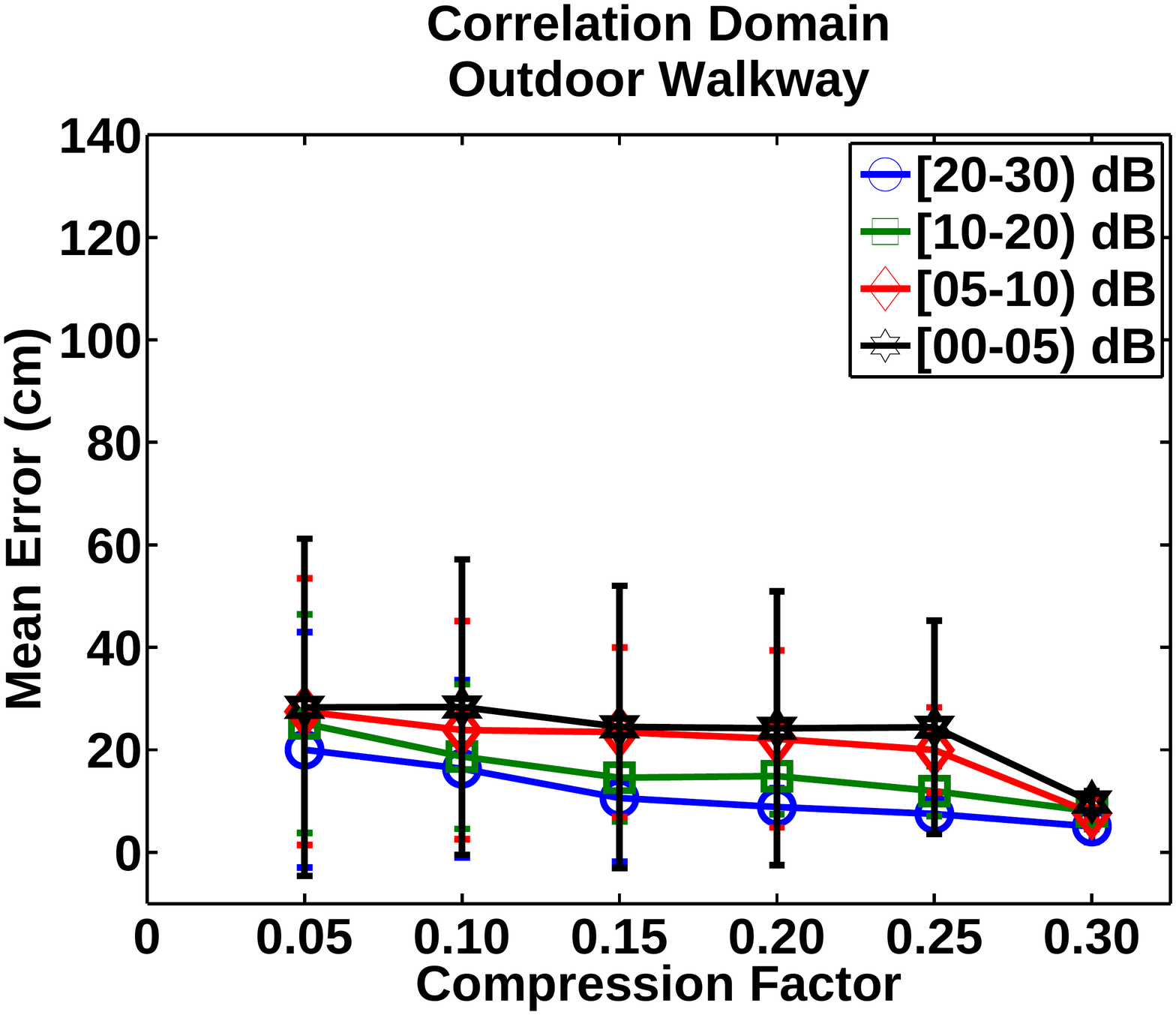} & \includegraphics[width=2.25in]{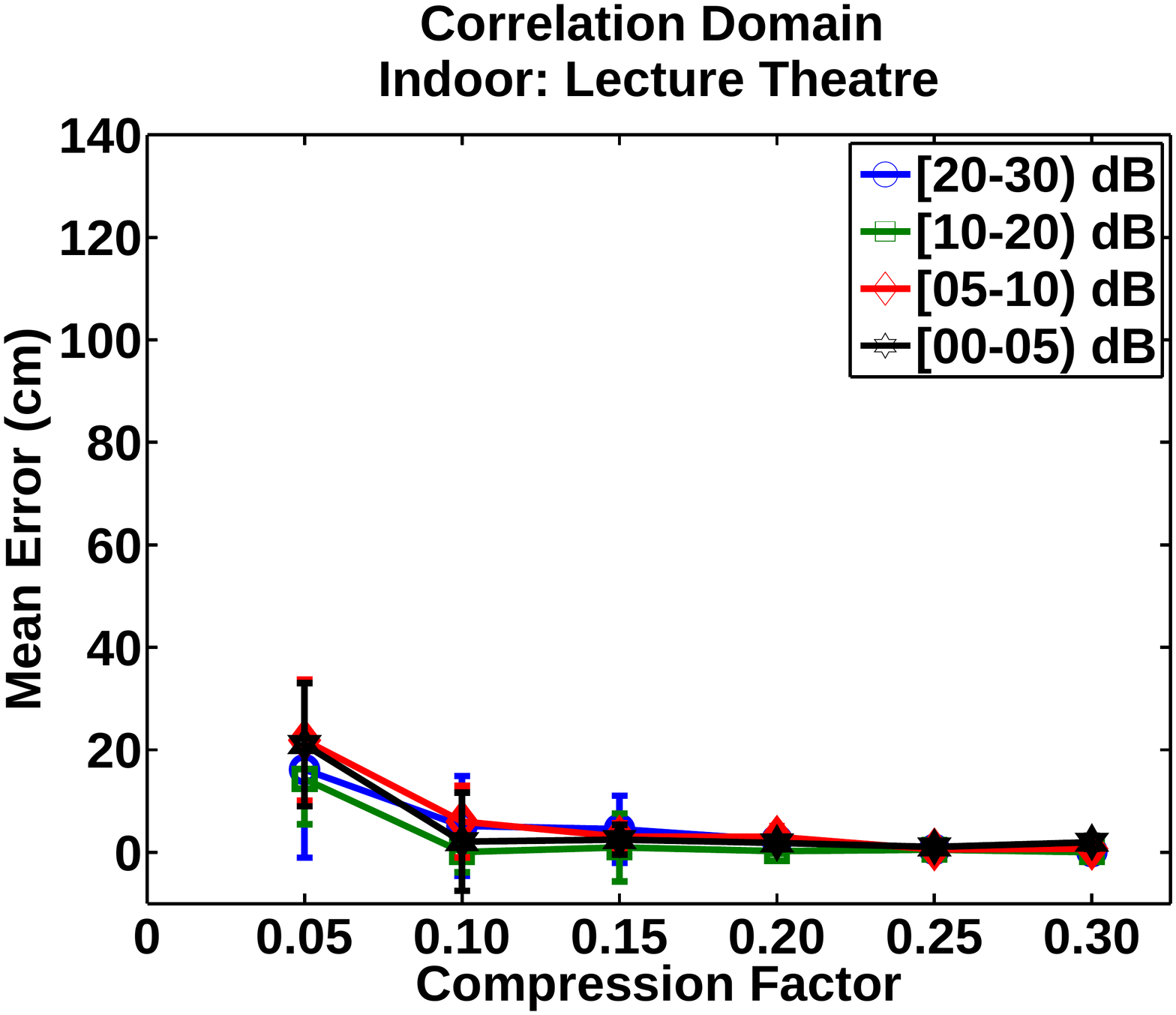} & \includegraphics[width=2.25in]{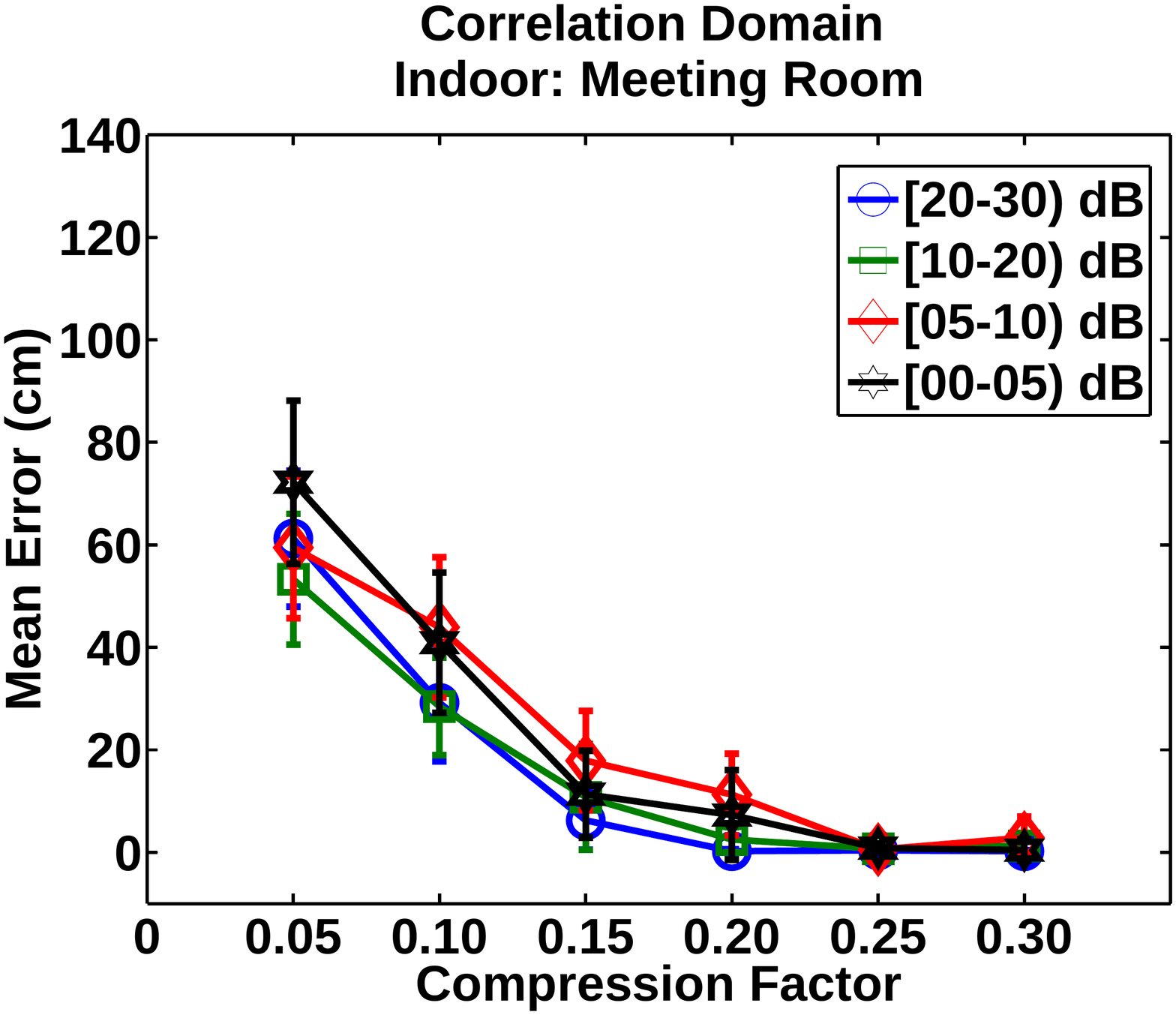}\\
(a) Case-A:  & (b) Case-B: & (c) Case-C: \\
Multipath: very-low & Multipath: low & Multipath: high\\
\end{tabular}
\end{center}
\vspace{-3mm}
\caption{\textbf{StructS-XCorr.} \emph{Characterization of compression factor $\alpha$ with SNR.}}
\label{fig:StructSparseXCorr}
\end{figure*}
\subsubsection{Characterization Studies and Benchmarks} \label{sec: characterization_studies}
\vspace{1mm}
\noindent
\textbf{Ranging error vs. \{compression factor, SNR\}.}
The optimal choice of the compression factor $\alpha$ that achieves the best accuracy with the least measurements (or projections) $m$ is a key design decision as a smaller $m$ leads to lower storage and transmission cost.
$\alpha$ depends on the sparsity $k$ (Eq.~\ref{eq:8}) of the received signal in the correlation domain, which in turn depends on the received SNR that varies with transmission power and ranging distance. 
In this subsection, we empirically study the relationship between SNR and $\alpha$.
The study was conducted in the following environments.
\vspace{1mm}
\newline
\noindent
$\bullet$ \textbf{Case-A} \emph{\{outdoor, very low multipath\}}: A less frequently used urban walkway, and the weather being sunny with occasional mild breeze.
\vspace{1mm}
\newline
\noindent
$\bullet$ \textbf{Case-B} \emph{\{indoor, low multipath\}}: A quiet lecture theatre ($[25\times15\times10]$\,m) with a spacious podium at one end of the large room.
\vspace{1mm}
\newline
\noindent	
$\bullet$ \textbf{Case-C} \emph{\{indoor, high multipath\}}: A quiet meeting room ($[7\times6\times6]$\,m) with a big wooden table in the center and other office furnitures.
\newline
\noindent
The transmitter and the receiver were fixed at a constant separation distance of $5$\,m.
The transmit power was varied such that the received SNR were recorded within the limits: $[0$-$5)$\,dB, $[5$-$10)$\,dB, $[10$-$20)$\,dB, $[20$-$30)$\,dB.
For reasons that will be explained in the next subsection, we slightly modified the peak selection criteria of the detection algorithm to choose the tallest peak if there was no valid peak ($6$ standard deviation above the mean).   
100 observations were collected for every experiment. 
\textit{We show the relative mean error and its deviation with respect to the (best-case) XCorr in all the results in this segment.}
\newline
\indent
Fig.~\ref{fig:fixed_dis_variable_power}(a), Fig.~\ref{fig:fixed_dis_variable_power}(b) and Fig.~\ref{fig:fixed_dis_variable_power}(c) shows the dependence of $\alpha$-compression and its recovery accuracy on the SNR of the ranging signal using \emph{S-XCorr}.
Across all figures, we observe that applying a higher $\alpha$ on a lower SNR signal results in an increase in estimation error.
Fig.~\ref{fig:fixed_dis_variable_power}(a) for Case-A presents the most clear characterization by negating the effect of channel multipaths (though introducing an increased background noise level), where observations with a high SNR of $[20$-$30)$\,dB provide reliable range estimates by using only $15$\% projections while those having low SNR of $[0$-$5)$\,dB show confident result only with $\alpha = 0.30$ (i.e., using more projections). 
Fig.~\ref{fig:fixed_dis_variable_power}(b) and Fig.~\ref{fig:fixed_dis_variable_power}(c) show the results for Case-B and Case-C.
Due to a less dominant multipath profile and background noise in Case-B, the accuracy levels show high confidence for $\alpha \geq 0.20$.
The situation is challenging in Case-C (due to high multipath), and so, the errors are as large as $1$\,m with $\alpha = 0.05$, but attain stability after $\alpha = 0.25$. 
The cumulative probability results suggest that there is a $95$\% probability of incurring an additional error of $< 1.5$\,cm in indoors and $< 3$\,cm in outdoors with $\alpha = 0.30$ with respect to its \emph{XCorr} estimate.
Using $\alpha > 0.30$ does not improve the accuracy significantly considering the additional overheads.
Fig.~\ref{fig:fixed_dis_variable_power} also shows that for applications that require lower accuracy (e.g., $100$\,cm), $\alpha$ as less as $0.05$ is sufficient.
We also performed ranging experiments with changes in distance over $1$-$10$\,m.
Although, smaller values of $\alpha$ (i.e., lesser projections) were good for high SNR levels, the results with $\alpha = 0.30$ were optimal, even in the worst case to obtain higher accuracy ($< 2$\,cm).
\begin{figure*}[t]
\begin{center}
\begin{tabular}{ccc}
\includegraphics[width=2.25in]{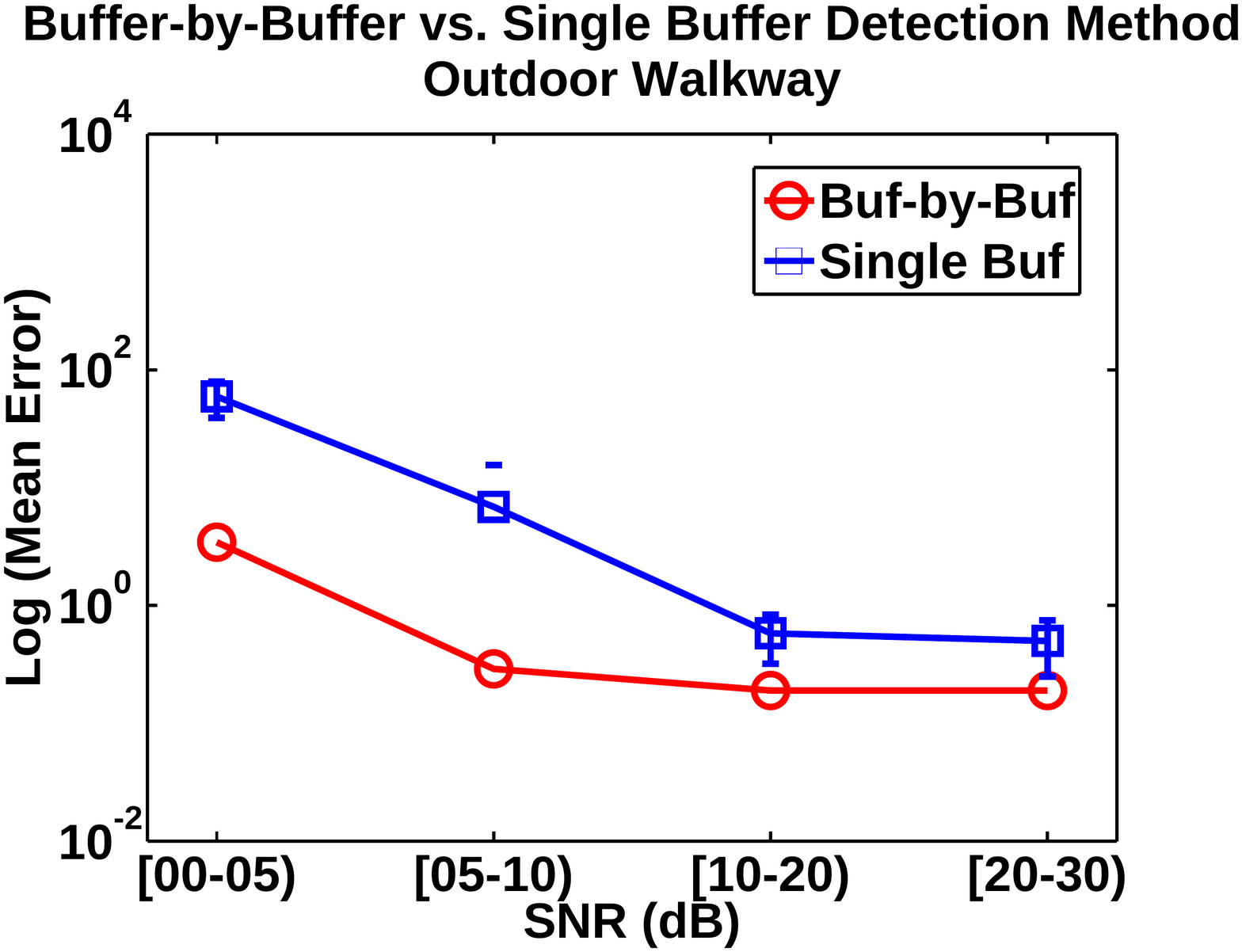} & \includegraphics[width=2.25in]{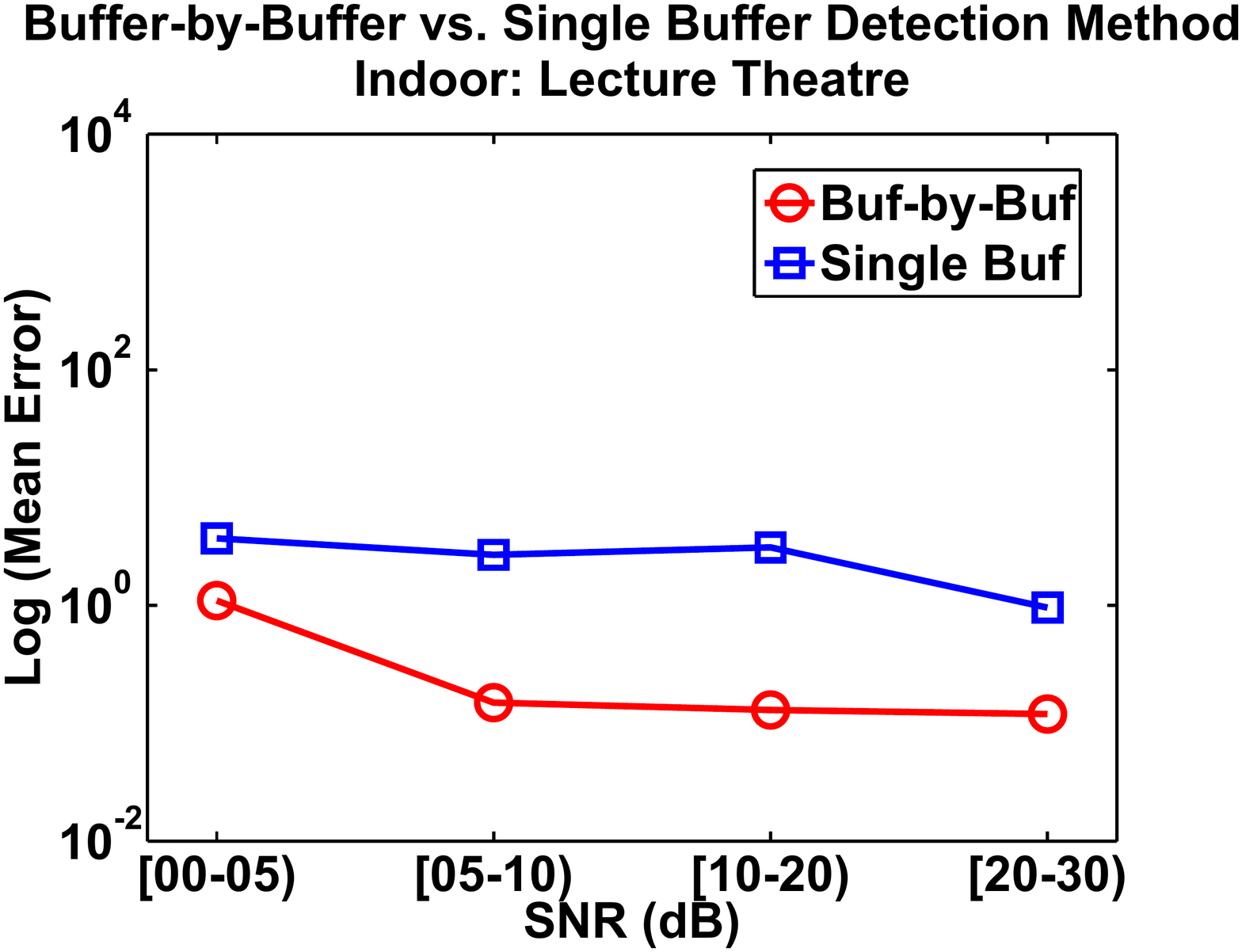} & \includegraphics[width=2.25in]{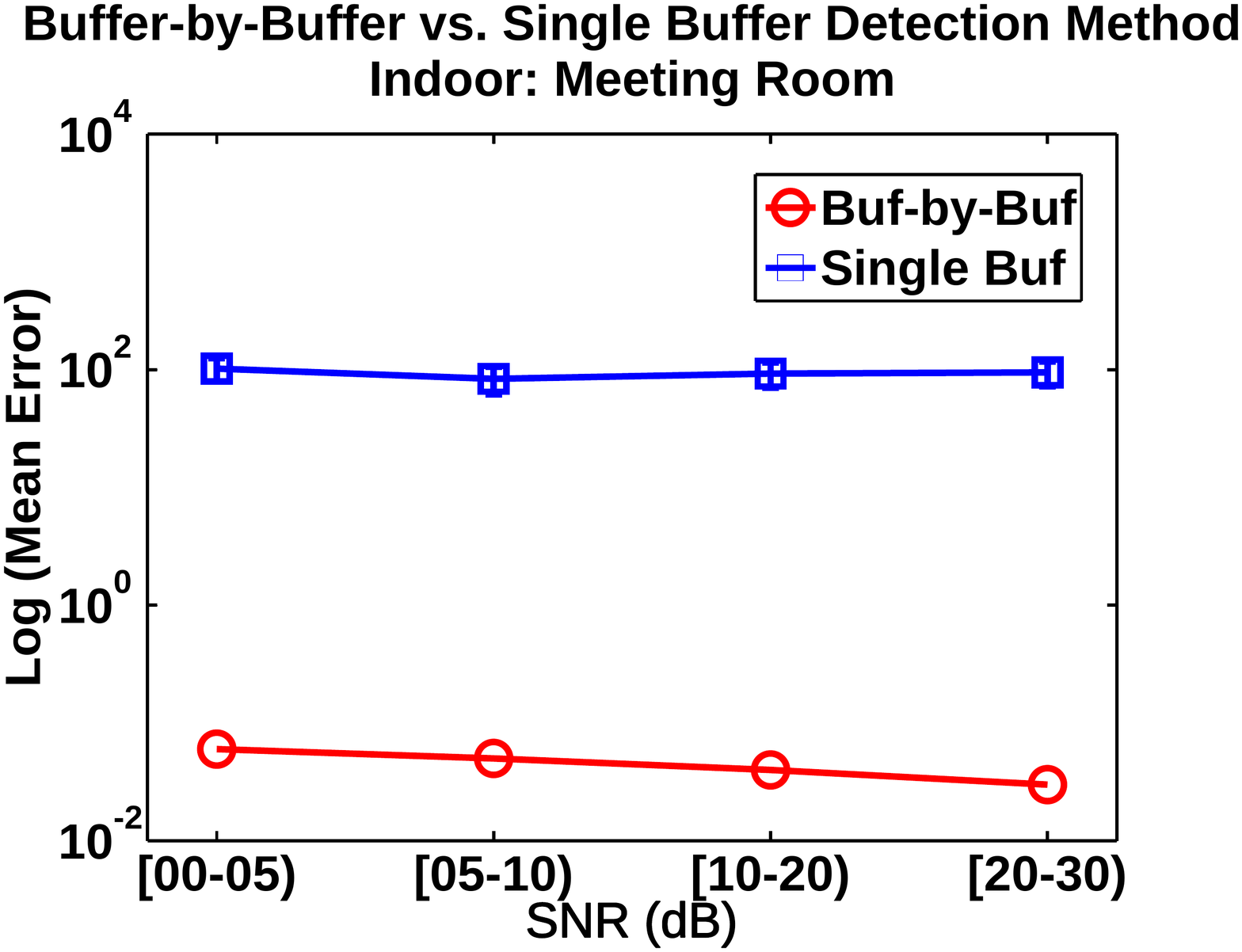}\\
(a) Improvement: & (b) Improvement:  & (c) Improvement: \\
Order of mag.~$1$ & Order of mag.~$1.5$ & Order of mag.~$4$\\
\end{tabular}
\end{center}
\caption{\textbf{StructS-XCorr: Buffer-by-Buffer vs. Single Buffer Detection.} \emph{For a compression factor of $0.30$, the buffer-by-buffer detection shows an order of magnitude $1$-$4$ improvement over single buffer detection method.}}
\label{fig:buf_comp}
\end{figure*}
\begin{figure*}[t]
\begin{center}
\begin{tabular}{ccc}
\includegraphics[width=2.25in]{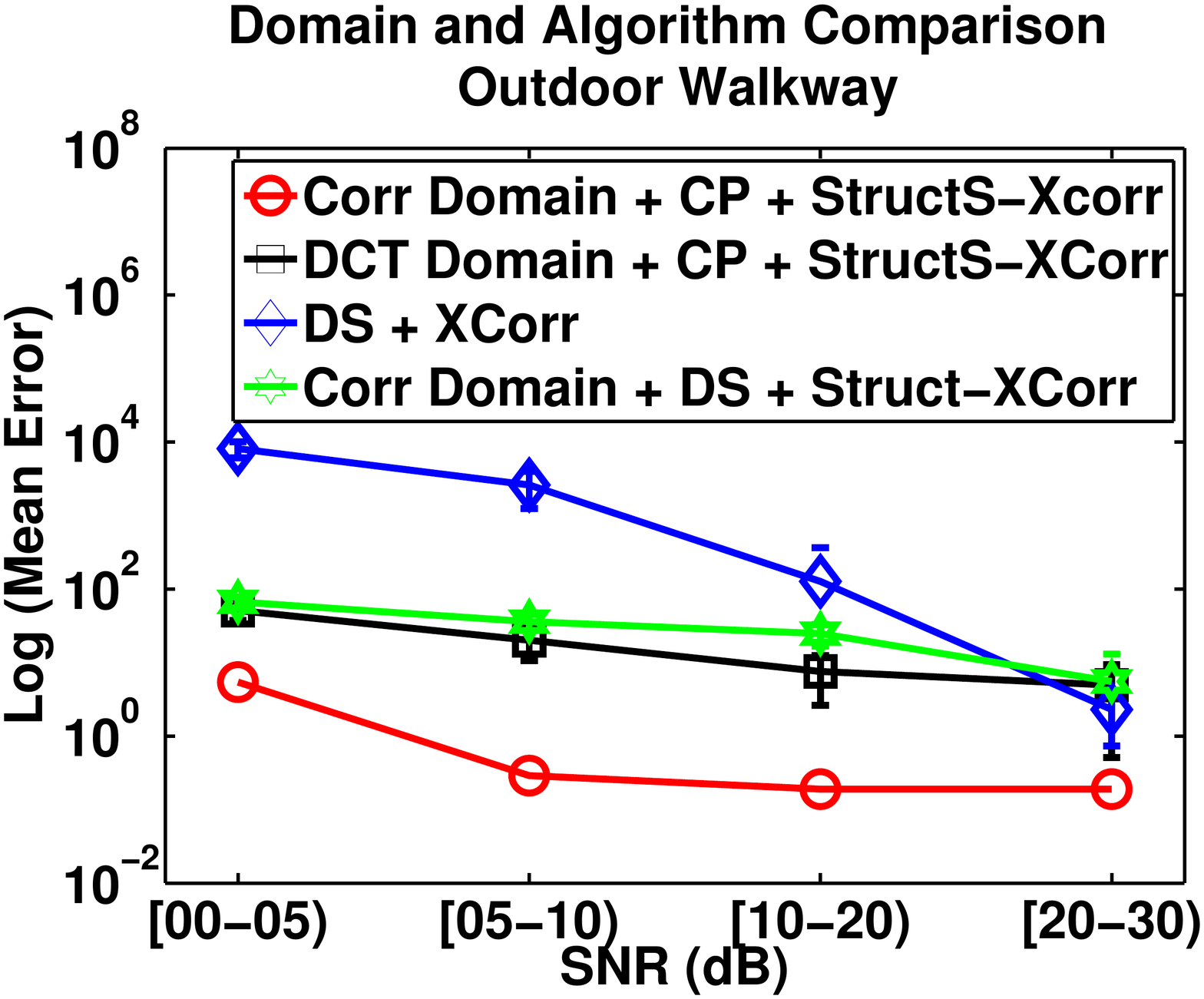} & \includegraphics[width=2.25in]{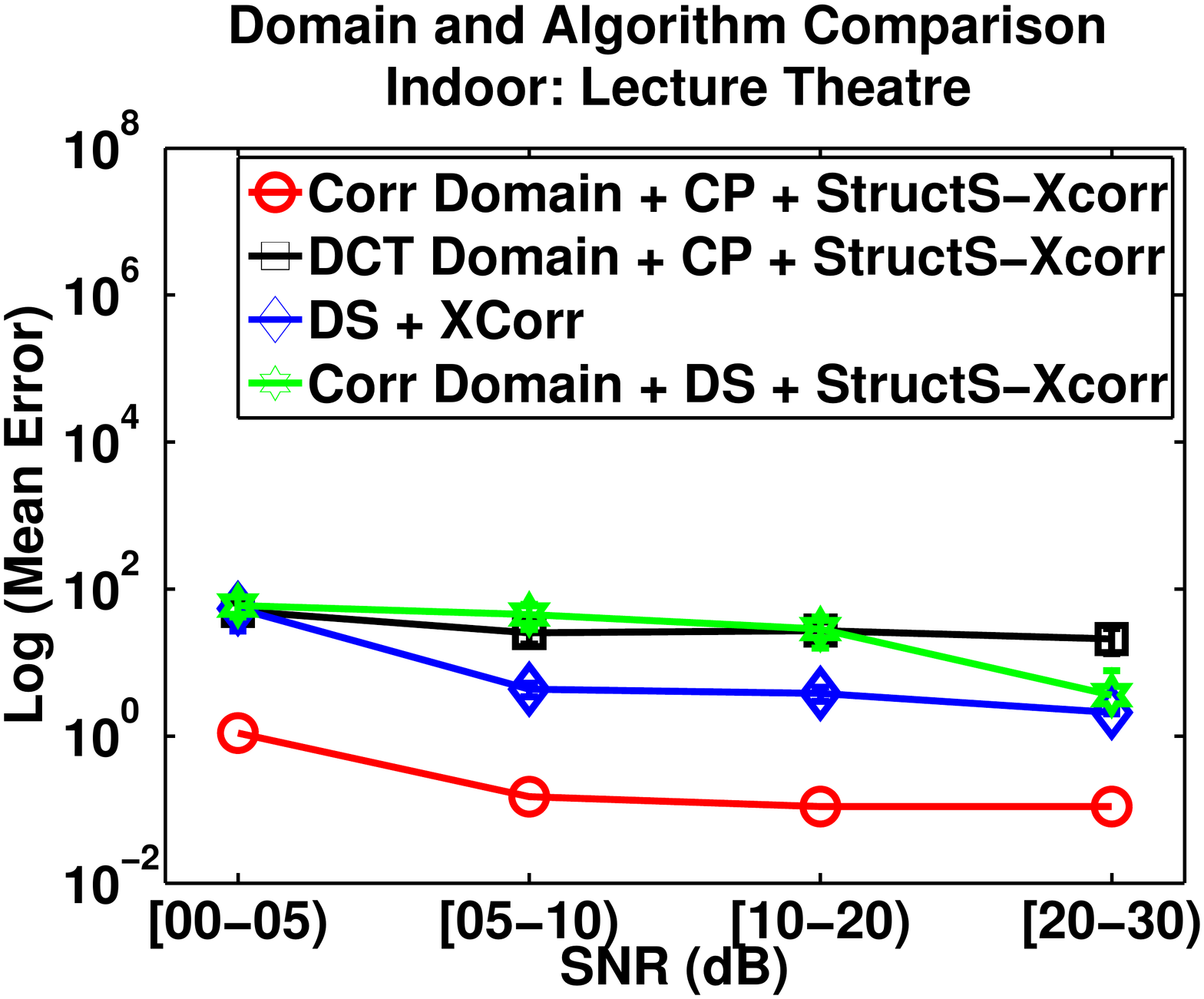} & \includegraphics[width=2.25in]{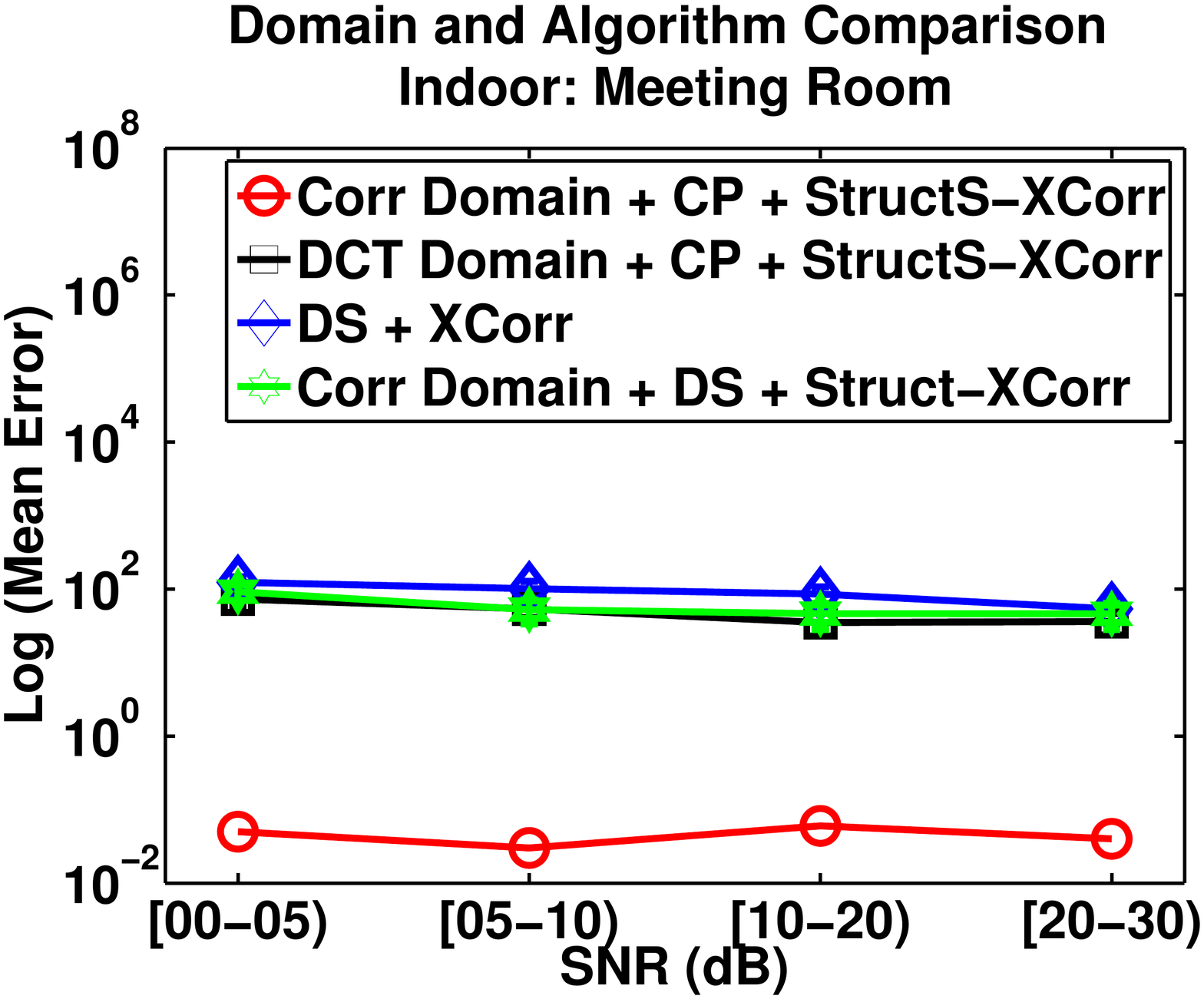}\\
(a) Improvement: & (b) Improvement:  & (c) Improvement \\
Order of mag.~$1$-$4$ & Order of mag.~$2$-$4$ & Order of mag.~$4$\\
\end{tabular}
\end{center}
\vspace{-1mm}
\caption{\textbf{Domain and Algorithm Comparison.} \emph{For a compression(CP)/downsampling(DS) factor of $0.30$, StructS-XCorr model in the correlation domain shows an order of magnitude $1$-$4$ higher detection accuracy compared to: (i) compression with StructS-Xcorr in the DCT domain (ii) downsampling with XCorr (iii) downsampling with StructS-XCorr.}}
\label{fig:algo_comp}
\end{figure*}
\newline
\indent
Fig.~\ref{fig:StructSparseXCorr} shows the above analysis, but using \emph{StructS-XCorr}.
It revelas \emph{two} interesting observations.
\emph{First}, similar to Fig.~\ref{fig:fixed_dis_variable_power}, we observe that applying a higher $\alpha$ on a lower SNR signal results in an increase in estimation error.
\emph{Second} and more imporantly, there is a $40$\% improvement in ranging accuracy for cases of high compression factor and low SNR in Case-A and Case-C.
However, there isn't appreciable improvement in Case-B (an environment with low noise).
\newline
\indent
Fig.~\ref{fig:buf_comp} compares the detection accuracy between our proposed buffer-by-buffer method versus processing all the samples in a single buffer using the \emph{StructS-XCorr} recovery method.
From reasons explained in Section~\ref{sec:detection}, the results show at least $1$ order of magnitude improvement.
\newline
\indent
The sparse representation in the proposed correlation domain shows significantly better accuracy of an order of magnitude $2$ (Fig.~\ref{fig:algo_comp}) compared to the DCT domain (for $\alpha$=$0.30$) due to the most sparse depiction of the ranging signal (Fig.~\ref{fig:sparsity}).
For DCT domain processing, the recovered coefficients $\hat{\textbf{s}}_{1}$ were multiplied with the DCT basis $\Psi$ (Eq.~\ref{eq:2}) to obtain an estimate of the received signal $\hat{\textbf{x}}_{1}$, and then cross-correlated with the reference signal $\textbf{p}$.
Here also, the recovery mechanism is based on \emph{StructS-XCorr}.
\newline
\indent
Another simple (but deterministic) method of reducing the sample count is to downsample \textbf{x} by a factor $F_d$ resulting in $\hat{\textbf{y}}$.
We verify its detection accuracy in the correlation domain by using two different algorithms: (a) standard cross-correlation and (b) \emph{StructS-XCorr} with the following formulation of the sparse approximation problem:
\begin{equation}
\vspace{-1mm}
	(\ell^{1}_{r}): \hspace{0.2cm} \hat{\textbf{s}}_{1}^{d} = \min \| \textbf{s} \|_{\ell_{1}} \hspace{0.2cm} \mbox{subject to:} ||\Psi^{'} \textbf{s} - \hat{\textbf{y}}||_{2} \leq \epsilon 
	\vspace{-1mm}
\end{equation}
\vspace{-4mm}
\newline
\noindent
The comparison results in Fig.~\ref{fig:algo_comp} show that neither of these two methods based on downsampling provide better estimates than the proposed method of $\ell^{1}$-minimization and structured sparsity in the correlation domain where the improvement is of an order of magnitude $2$ across all experimental environments.
Information embedding in random ensembles preserves the $\ell^{2}$-norm (or energy) of its respective higher dimension representation, and therefore, the recovery accuracy is significantly better than deterministically choosing samples and discarding information (i.e., frequency components) by downsampling.
This result, therefore, supports the theoretical underpinning that there is an overwhelming probability of correct recovery via $\ell^{1}$-minimization for dimensionality reduction by random linear projection (Section~\ref{sec:theory}).
Since the recovery techniques are based in l1-minimization, we direct the readers to \cite{fastl1} for a systematic benchmark of their performance.
\vspace{1mm}
\newline
\noindent
\textbf{Adaptive estimation: compression factor.}
The design of an adaptive mechanism for $\alpha$ requires estimating the received SNR.
We propose two different approaches: first, with a BS feedback to receiver, and second, on the receiver itself.
\newline
\indent
For the BS-feedback mechanism, we utilize empirical information from the peak detection algorithm.
In Section~\ref{sec:detection}, we considered the scenarios where the valid buffer count $\tilde{b} \geq 1$.
If a valid peak (i.e., at least $6$ standard deviations above the mean) is not detected in any buffer (i.e., $\tilde{b}$ = $0$), then the detection is considered to have failed.
This implies that the recovered coefficients are noisy due to a non-optimal $\alpha$ for the respective measurements (characterized by its SNR).
It was precisely the reason for modifying the peak selection criteria in the previous subsection, where we observed large errors in peak positions for magnitudes below the specified threshold.
The BS-feedback algorithm starts with the initial knowledge of whether a valid peak was determined with $\alpha$ = $0.30$.
If the detection succeeds, then $\alpha$ is decremented by a step size of $0.05$ and compressed.
This process is iterated until the detection fails, in which case, the previous $\alpha$ values is selected.
On the other hand, if no valid peaks were encountered for the starting case, $\alpha$ is incremented in steps of $0.05$ and the entire process is repeated until the detection succeeds.
\newline
\indent
A major drawback of the feedback approach is the additional measurements (that translate to transmission overhead), and its associated delay and power usage for deriving $\alpha$.
Therefore, we introduce this functionality on the receiver by a simple power estimation algorithm.
The ratio $\rho$ of the peak signal amplitude to the average of the absolute values in the sampled signal is calculated, and a corresponding $\alpha$ is selected according to the following empirically chosen criteria. 
$\alpha = \{\{0.05: \rho > 30\},\{0.10: 20 < \rho \leq 30\},\{0.10: 20 < \rho \leq 30\},\{0.20: 15 < \rho \leq 20\},\{0.30: 10 < \rho \leq 15\},\{0.50: 05 < \rho \leq 10\},\{1.00: \rho \leq 05\}\}$.
\newline
\indent
For our analysis, we randomly selected $1000$ measurements pertaining to different SNR levels in the indoor lecture theatre (Case-B).
The respective $\alpha$ was estimated using the above two methods and their performance was compared against our empirically selected threshold value of $\alpha$ = $0.30$.
Table \ref{tab:tradeoff_adaptive} reports their performance trade-off where the BS-feedback obtains high accuracy but requires $2$~times more measurements, while the receiver estimation approach takes fewer measurements and obtains only a $5$\% worse accuracy.
\begin{table}[t]
\begin{center}
\caption{\textbf{Projections vs. Accuracy.} \emph{A positive value indicates higher projections or reconstruction error compared to the threshold $\alpha$ = 0.30}}
\begin{small}
\begin{tabular}{lcc}
\toprule
	\textbf{Scenario} & \textbf{Projections (\%) } & \textbf{Accuracy (\%)} \\
	\midrule
	BS-Feedback &  \hspace{1mm} 101.16 &  -1.75 \\
	Receiver &  -17.55 &  \hspace{1mm}5.26 \\
\bottomrule
\label{tab:tradeoff_adaptive}
\vspace{-10mm}
\end{tabular}
\end{small}
\end{center}
\end{table}


\section{Evaluation} \label{sec:evaluation}
We present the design and implementation of an end-to-end acoustic ranging system using constrained WSN platforms in this section, followed by evaluation results.
\emph{Fast} data acquisition and compression on the receiver node was the underlying system rationale; hence, all design decisions were guided towards maximum RAM utilization rather than external flash (that would introduce additional latency).

\begin{figure}[b]
\begin{center}
\begin{tabular}{c}
\includegraphics[width=3.4in]{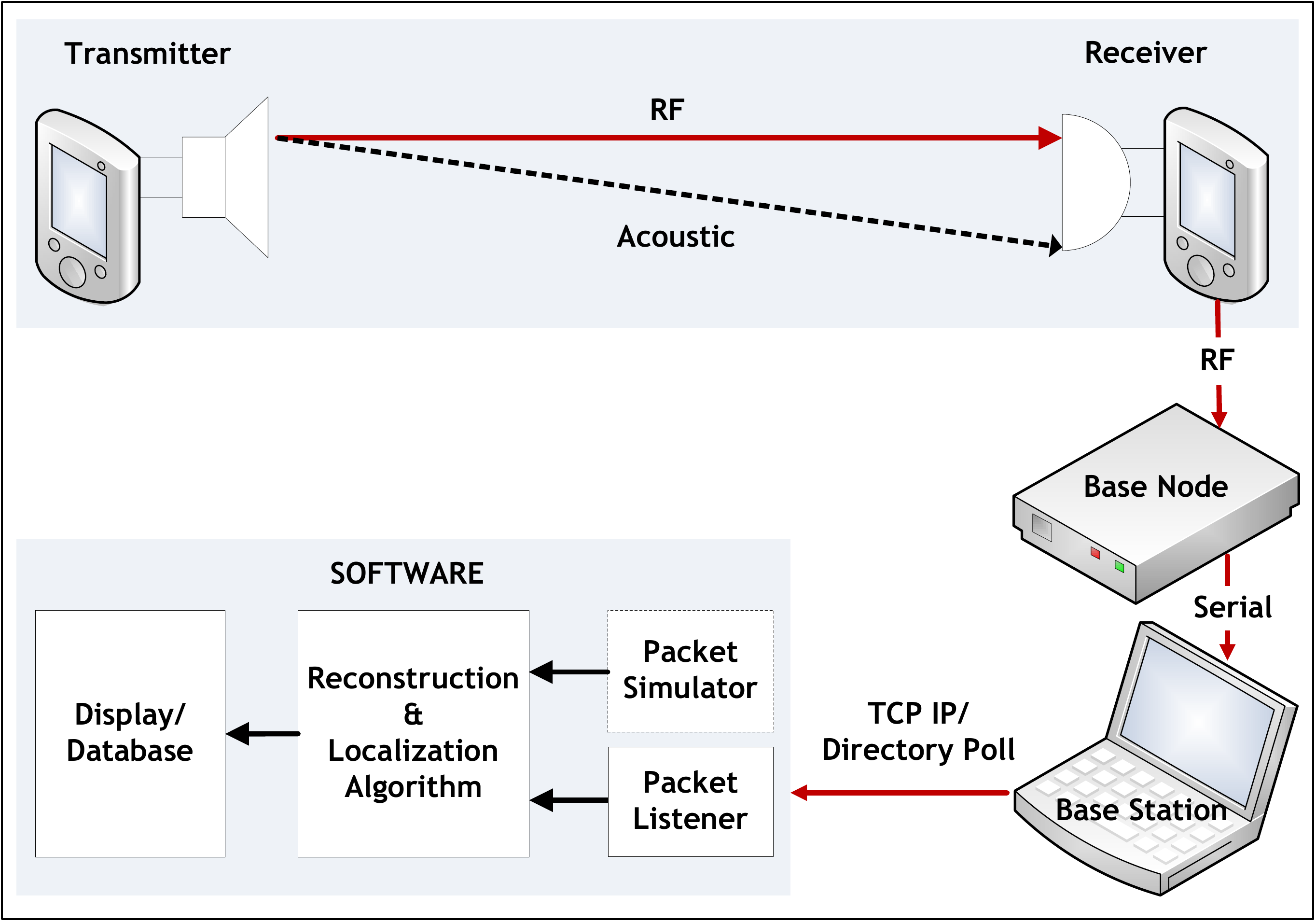} \\
\end{tabular}
\end{center}
\vspace{-4mm}
\caption{\textbf{System architecture.} \emph{End-to-end acoustic ranging system using constrained WSN platforms.}}
\label{fig:compressive_ranger_sys_model}
\vspace{-6mm}
\end{figure}

\subsection{System design on constrained platforms.}
\vspace{1mm}
\noindent
The system comprised of the TmoteInvent (as listener), our designed sensor mote (as beacon) and a network interface to the base-station (Fig.~\ref{fig:compressive_ranger_sys_model}).
\vspace{1mm}
\newline
\noindent
\textbf{Transmitter.} 
The beacon node \cite{Misra2013:acoustic_bible} comprised of our WSN platform along with a custom designed audio daughter board that included four TI TLV$320$AIC$3254$ audio codecs and the Bluetechnix CM-BF$537$E digital signal processor module. 
The transmitting front-end of the beacon mote consisted of a power amplifier driving a tweeter (speaker) transducer (VIFA $3/4$'' tweeter module MICRO).
The tweeter (size: [$2\times2\times1$]\,cm) had a fairly uniform and high frequency response of $\approx 22$\,dB above the noise-floor between $1$-$10$kHz.\vspace{1mm}
\newline
\noindent
\textbf{Receiver.}
TmoteInvent\cite{tmoteinvent} was used as the listener node, due to its low-cost and low-power ($100$ times more power efficient than the DSP on the transmitter) features that are expected from a WSN platform. 
The receiving front-end consisted of an omni-directional electret microphone (Panasonic WM-$61$B) attached to an Analog Devices SSM$2167$ preamplifier.
It allows omni-directional acquisition in the range $20$\,Hz - $10$\,kHz, and has a near-flat frequency response between $3$-$7$\,kHz that is $10$\,dB above the noise floor.
High-rate audio data collection was achieved using the DMA controller packaged with the MSP$430$ MCU.
However, the MSP$430$ DMA causes truncation of the $12$\,bits ADC data to 8 bits rather than to two bytes, and so, results in a data resolution loss of $4$\,bits.  
\begin{figure*}[t]
\begin{center}
\begin{tabular}{ccc}
\includegraphics[width=2.25in]{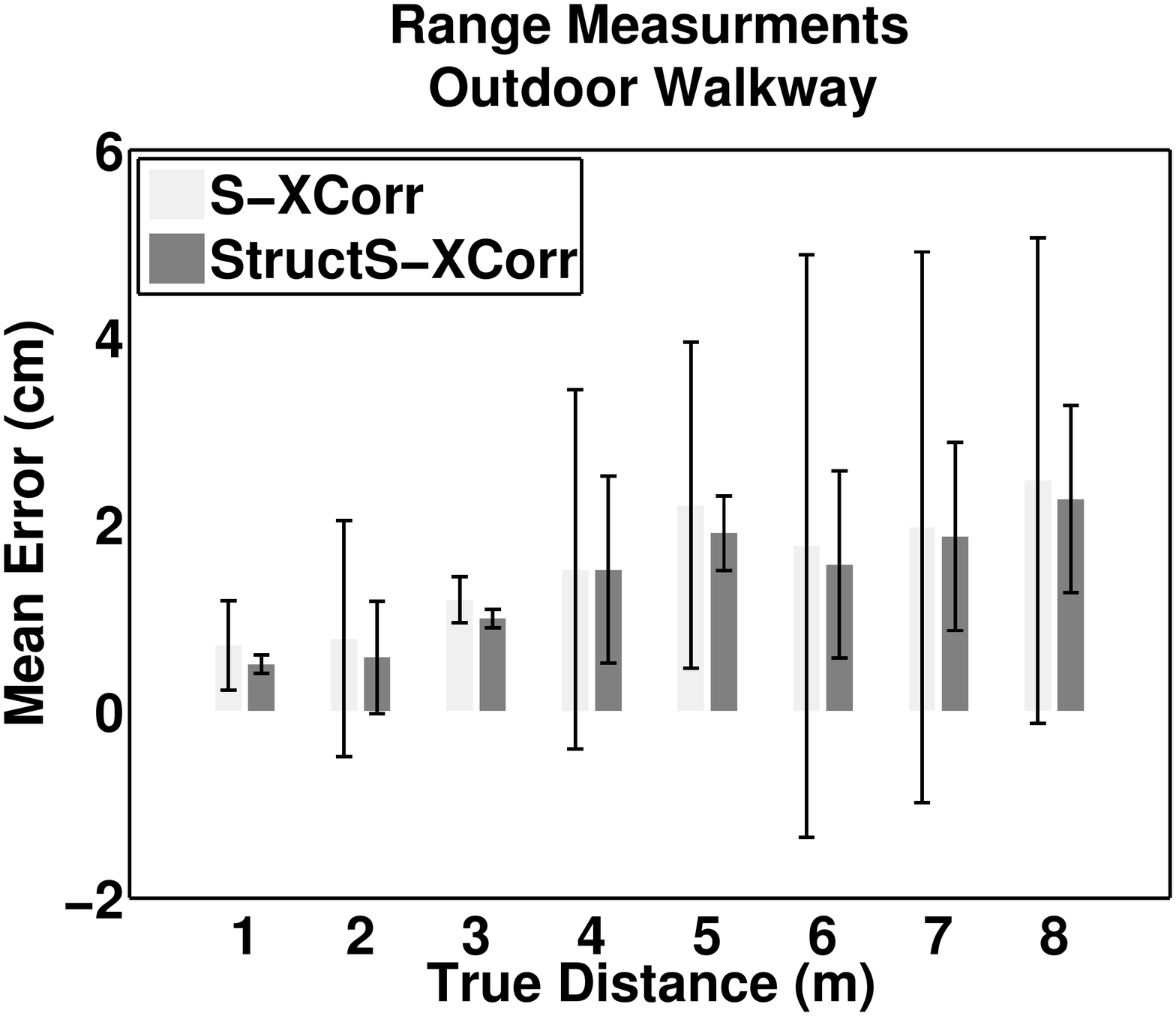} & \includegraphics[width=2.25in]{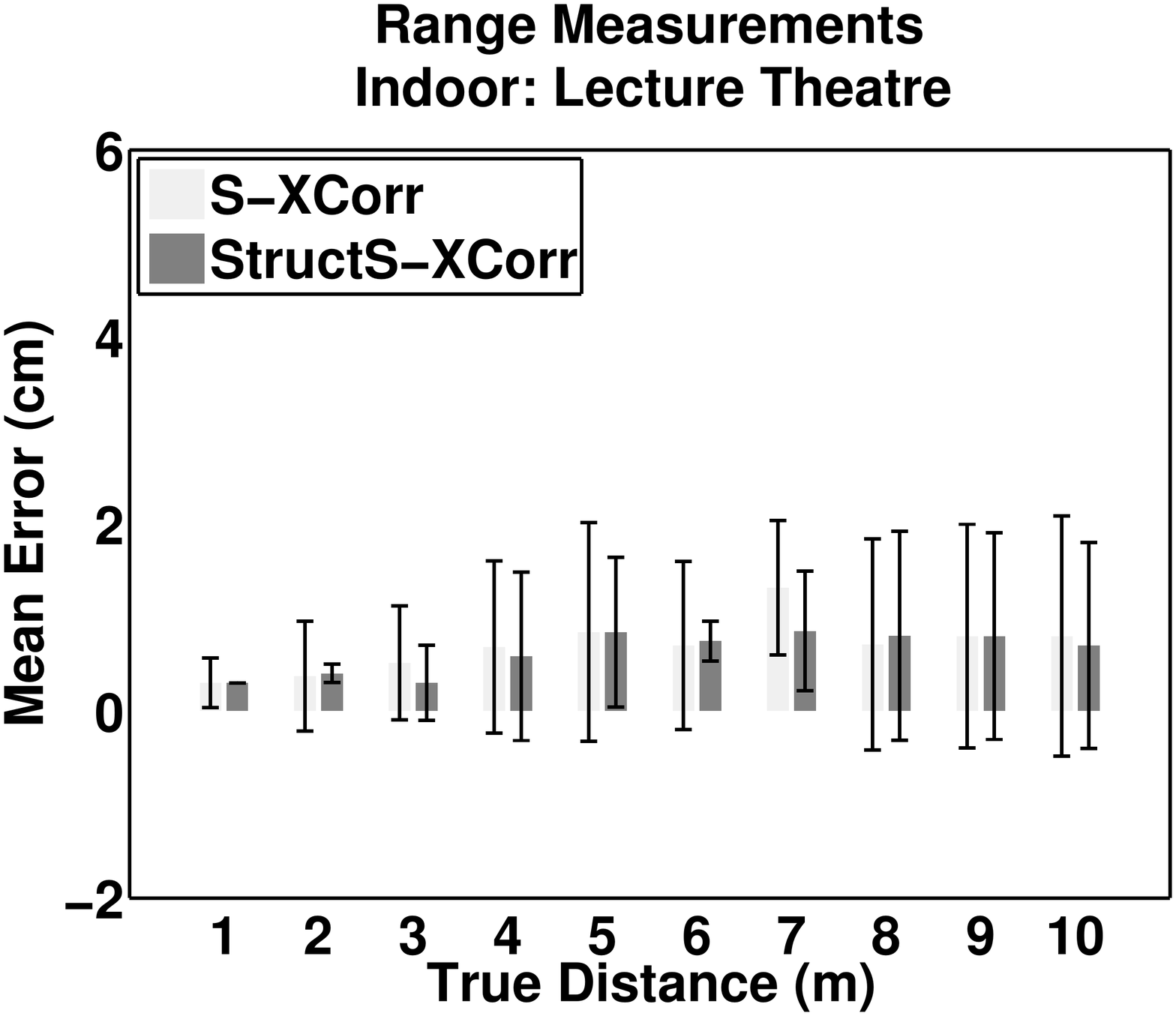} & \includegraphics[width=2.25in]{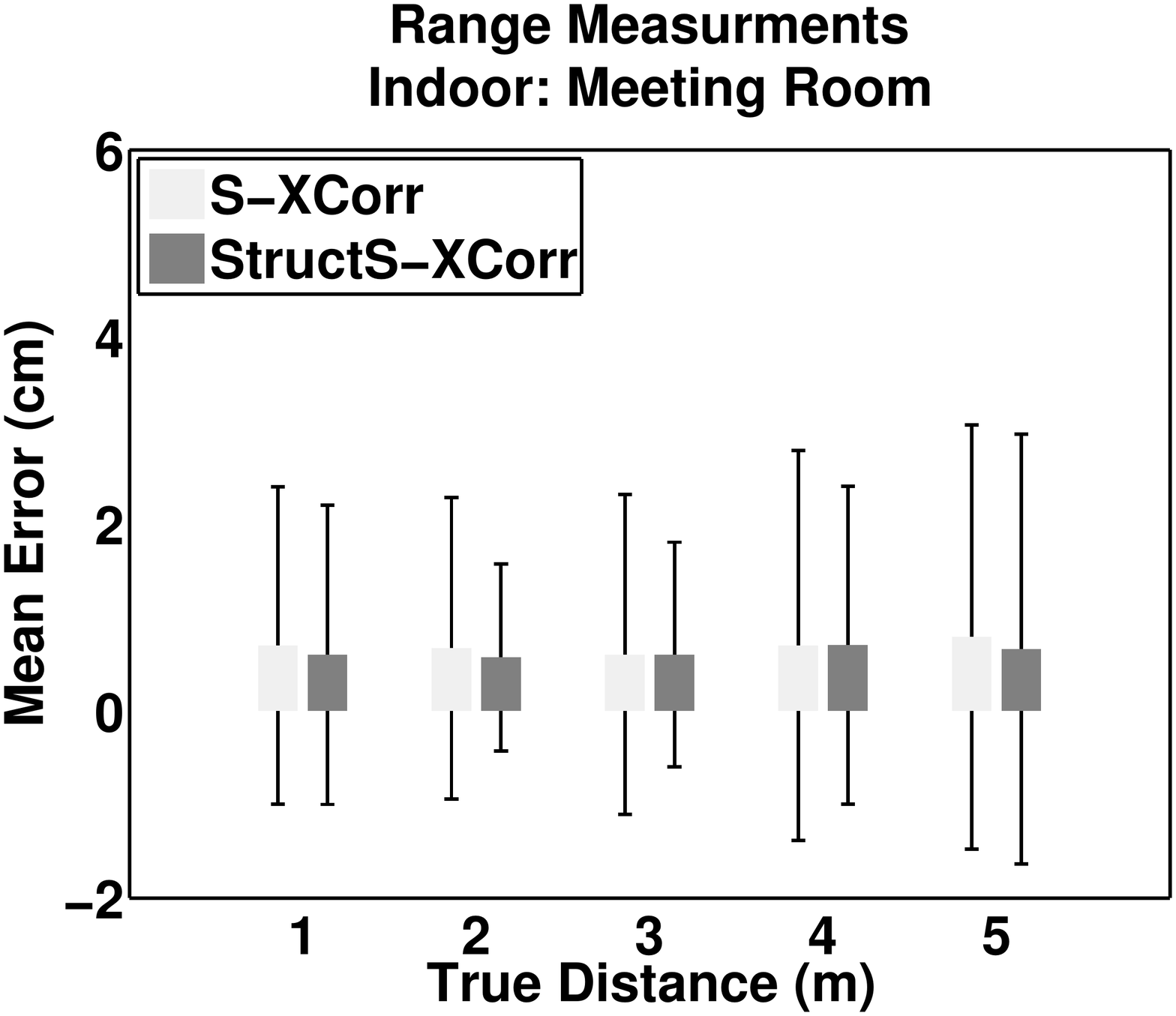} \\
(a) Case-A: & (b) Case-B: & (c) Case-C: \\
Very-low Multipath & Low Multipath & High Multipath \\
$\alpha$ = \{$0.30$ ($1$-$5$\,m)\} & $\alpha$ = \{$0.30$ ($1$-$5$\,m)\} & $\alpha$ = \{$0.30$\}\\
$\alpha$ = \{$0.40$ ($6$-$8$\,m)\} & $\alpha$ = \{$0.35$ ($6$-$10$\,m)\} & \\
\end{tabular}
\end{center}
\vspace{-2mm}
\caption{\textbf{End-to-End acoustic ranging system using constrained WSN platforms.} \emph{Ranging results.}} 
\label{fig:evaluation_results}
\vspace{-4mm}
\end{figure*}
\newline
\noindent
\textbf{Ranging/Detection methodology.}
The system uses the V-TDOA of RF and acoustic signals to measure the beacon-to-listener distance.
The beacon initiates the ranging process by periodically transmitting a RF signal followed by a acoustic pulse after a fixed time interval.
The fast propagating RF pulse reaches the listener almost instantaneously and synchronizes the clocks on both the devices, following which, the TDOA is measured after the arrival of the acoustic pulse.
The ranging signal was a linear chirp of [$3$-$7$]\,kHz/$0.01$\,ms and was transmitted at an acoustic pressure level of $70$\,dB.
The DAC on the audio codec of the beacon node was programmed to sample at $48$\,kHz, while the ADC on the receiver Tmote was configured to acquire at $15$\,kHz.
\newline
\indent
If the time taken for sound to travel a maximum range $d_{c}$ at a speed $v_s$ is at most $\frac{d_{c}}{v_s}$, and if the transmitted chirp length is $t_{p}$, then the signal must reach the receiver within [$\frac{d_{c}}{v_{s}} + t_{p}$].
For $t_{p}$ = $0.01$\,s  and $d_{c} \approx$ $10$\,m, the recording of the signal must be completed by $0.03$\,s.
We include an additional $0.01$\,s to compensate for reverberation time ($t_{c}$), and setup the recording time to $0.04$\,s (Eq.~\ref{eq:acqtime}).
Following the buffer-by-buffer compression method, the signal was spread across $5$ buffers.
A measurement matrix $\bar{\Phi}$ was stored in the RAM that contained i.i.d. entries sampled from a symmetric Bernoulli distribution (Eq.~\ref{eq:11}).
We postponed the multiplication operation on the matrix entities with the constant ($1/\sqrt{m}$) until the recovery stage at the BS.
\newline
\indent
The listener acquires the audio samples, compresses and stores these measurements in the RAM over a period of $5$ iterations, and then, transfers them to the BS.
These measurements are again divided into their respective buffers and reconstructed to obtain the coefficients.
The detection process is the same as explained in Section \ref{sec:detection}, however we made two minor modifications.
First, due to a higher receiver noise floor, we set the criteria for selecting the first tallest peak to $3$ (instead of $6$) standard deviations above the mean.
Second, as each sample corresponds to $2.2$\,cm of distance (at a sampling rate of $15$\,kHz), we used a simple parabolic interpolation method\cite{Misra2011:tweet} to obtain finer resolution.
This additional step identifies the position of the first neighboring peak on the left and right of the selected ranging peak, finds the parabola that
passes through these points, and calculates the time coordinate of the maximum of this parabola that estimates the range. 
\vspace{-2mm}
\subsection{Performance Results}
\noindent
\textbf{Ranging.}
The ranging experiments were performed in the same three environments as mentioned in Section \ref{sec: characterization_studies}: (a) Case-A: outdoor walkway, (b) Case-B: lecture theatre, and (c) Case-C: meeting room.
In all the setups, the listener node was fixed while the beacon node was moved along the direct LOS in a controlled manner.
The correct ground truth was established using a measuring tape and markers.
$30$ observations were collected for every experiment.
While the ranging performance of \emph{S-XCorr} in the absolute sense was studied in our previous work \cite{Misra2012:sparsexcorr}, here we study the relative performance of \emph{S-XCorr} and \emph{StructS-XCorr} with respect to the (best-case) \emph{X-Corr} in terms of the relative mean error and its deviation (Fig.~\ref{fig:evaluation_results}).
\newline
\indent
As shown in Fig.~\ref{fig:evaluation_results}(a), Case-A recorded the best result where the cumulative error (mean + deviation) of \emph{StructS-XCorr} with respect to \emph{S-XCorr} was approximately $1$\,cm for distance between $1$-$4$\,m; but improved by more than $2$\,cm for ranges beyond that.
Measurements after $5$\,m were compressed with an $\alpha = 0.40$.
The audio recordings after $8$\,m were highly noisy, and therefore, required an even higher $\alpha$ value for compression to compensate for the reduced sparsity levels.
However, due to non-availability of RAM memory space for storing the additional entries of the new measurement matrix $\bar{\Phi}$, range estimates beyond $8$\,m could not be processed.
Due to the decrease in the sparsity levels with lower SNR, the measurements from $[6-8]$\,m were compressed with a higher $\alpha$ of $0.40$.
Fig.~\ref{fig:evaluation_results}~(b) and Fig.~\ref{fig:evaluation_results}~(c) show the results for Case-A and Case-C, and it suggests that \emph{StructS-XCorr} did not significantly outperform \emph{S-XCorr}.
Fig.~\ref{fig:evaluation_results} supports the same observation that was noted in Fig.~\ref{fig:fixed_dis_variable_power} and Fig.~\ref{fig:StructSparseXCorr}, where \emph{StructS-XCorr} is beneficial in conditions of low SNR (typical of Case-A).
\vspace{1mm}
\noindent
\textbf{Localization.}
In these experiments, $5$ listener nodes were placed at fixed (known) locations in an outdoor setting (Case-A) to obtain the (unknown) location coordinates of the beacon node.
This layout was kept consistent with the approach presented in \cite{Misra2012:sparsexcorr}.
The speaker had a fairly uniform signal strength within the directionality cone of $\pm$ $45^{o}$ (with a $2$\,dB decrease from $0^{o}-45^{o}$), therefore, all the $5$ listeners were confined within this perimeter with their microphones facing the speaker. 
\newline
\indent
The beacon initiated the ranging process and the corresponding acoustic chirp was recorded by the $5$ listeners.
A simple time division multiple access (TDMA) approach was followed for orderly data transfer wherein each listener transferred the compressed data in a preset time slot.
The distances between the beacon and the receivers were estimated at the BS, which was followed by the linear least square localization algorithm\cite{Misra2013:acoustic_bible} to calculate the $2$D location of the beacon node.
Fig.~\ref{fig:localization} shows the node placements, where the listener and beacon node(s) have been depicted as circle and square respectively along with the estimated beacon location using the two methods.
The cumulative (mean + deviation) localization error between \emph{S-XCorr} and \emph{StructS-XCorr} was less than $1$\,cm.
As the localization error is upper bounded by its ranging errors, we expect similar relative performance in Case-B and Case-C that show a maximum ranging error difference of $2$\,cm.
\begin{figure}[t]
\begin{center}
\begin{tabular}{c}
\includegraphics[width=3.5in]{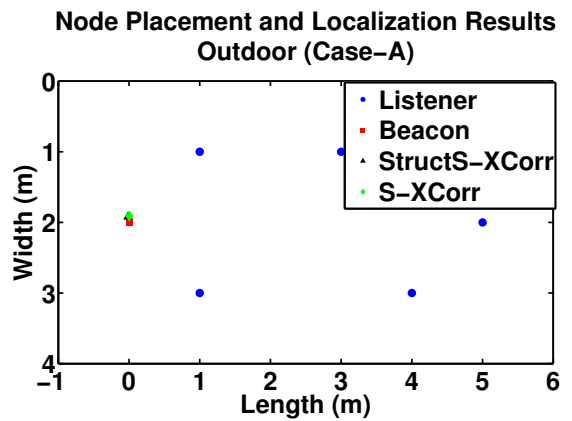} \\
\end{tabular}
\end{center}
\vspace{-1cm}
\caption{\textbf{End-to-End acoustic ranging system using constrained WSN platforms.} \emph{Localization results for Case-A.}} 
\label{fig:localization}
\end{figure}
\begin{table}[b]
	\caption{\textbf{Performance Analysis: TmoteInvent}}
	\label{tab:performance}
	\begin{center}	
	\begin{small}
	\begin{tabular}{lcc}
	\toprule
	\textbf{Operation} & \textbf{Time (s)} & \textbf{Energy (mJ)}\\ 
	\midrule
	Audio Acquisition & 00.0665 & 0020.50 \\ 
	\midrule
  Compression & 00.0060& 0001.85\\ 
  Radio Transfer (Compressed Data) & 00.0580 & 0017.88\\ 
  \midrule
  Cross-correlation (Time-domain) & 15.6250 & 4816.00 \\ 
  \bottomrule
	\end{tabular}
	\end{small}
	\end{center}
\end{table}

\begin{figure}[t]
\begin{center}
\begin{tabular}{c}
\includegraphics[width=3.5in]{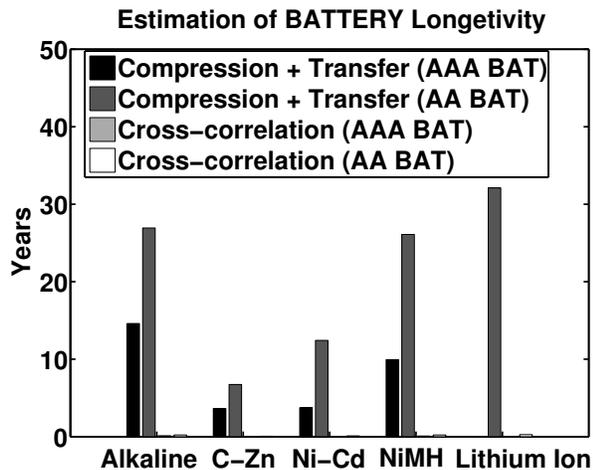} \\
\end{tabular}
\end{center}
\vspace{-0.5cm}
\caption{\textbf{Estimation of Battery Longetivity.} \emph{Compared to the method of executing the standalone cross-correlation algorithm on the receiver platform, the combination of compression (by random ensembles) followed by transmission to the base-station will extend the battery lifetime to multiple years.}} 
\label{fig:battery}
\vspace{-2mm}
\end{figure}

\begin{table}[b]
	\caption{\textbf{Compression Factor ($\alpha$) for LZ77-based Compression Algorithm `gzip'.} \emph{Dataset collected by the POC System (Section~\ref{sec:analysis-structs-xcorr}).}}
	\label{tab:gzip}
	\begin{center}	
	\begin{small}
	\begin{tabular}{lcc}
	\toprule
	\textbf{Scenario} & \textbf{Mean $\alpha$} & \textbf{Deviation $\alpha$}\\ 
	\midrule
  Case-A: Very-low Multipath & 0.27 & 0.005 \\ 
  Case-B: Low Multipath & 0.27 & 0.005 \\
  Case-C: High Multipath & 0.28 & 0.009 \\
  \bottomrule
	\end{tabular}
	\end{small}
	\vspace{-3mm}
  \end{center}
\end{table}
\vspace{2mm}
\noindent
\textbf{Energy Consumption.}
Table \ref{tab:performance} reports the time and energy consumed for each operational step on the listener node.
The cumulative time spent in compression and radio transfer is $\approx$ 0.0640\,s, which is more than $100$ times faster than performing time-domain cross-correlation on the node itself.
Its equivalent frequency-domain cross-correlation requires $2$*FFT and IFFT operation steps.
This translates to significant energy saving on the receiver device, wherein a typical AA or AAA battery can last for over a decade (Fig.~\ref{fig:battery}). 
When optimized for speed, a FFT over $512$ sample window of an $8$\,kHz signal takes $0.5$\,s execution time on TelosB \cite{Greenstein2006}, and so, for our case of $750$ samples would take $\approx 2.2$\,s, which is still $34$ times slower.
\newline
\indent
With respect to compression performance, the popular LZ$77$-based algorithm `gzip' achieves slightly better compressibility of $\alpha = 0.27$ (Table~\ref{tab:gzip}).
However, due to its lossless nature of compression, it is not robust to information loss (packet drops) that are common in low-power sensor networks.
In contrast, the performance degradation by our approach is less severe and has the same effect as compressing with a smaller $\alpha$ (Fig.~\ref{fig:fixed_dis_variable_power}).
A similar, but energy efficient algorithm proposed by Sadler et al. \cite{Sadler2006}: S-LZW reports an execution time of approximately $0.05$\,s for $528$ bytes of data, and therefore, its equivalent compressing cost for $750$ bytes would be approximately $0.075$\,s ($\approx 12$ times slower than our technique).
These statistics suggest that although compression by random ensembles is not the best compression method, it benefits of greater energy savings along with faster data processing is a good trade-off between compression and computation time, accuracy (in case of data loss), energy consumption. 
For example, if applications can tolerate $100$\,cm localization accuracy, the proposed method requires approximately $5$\% of measurements only (Fig.~\ref{fig:fixed_dis_variable_power}). 
Furthermore, in the event of packet loss, S-LZW needs to either retransmit the entire compressed data segment, or employ expensive end-to-end reliable communication protocol. 
On the other hand, the performance of the proposed protocol degrades gracefully with packet losses as it can still recover the ranging information, 
but with larger errors (Fig.~\ref{fig:StructSparseXCorr}).

\section{Related Work} \label{sec:related_work}
We broadly categorize the related work based on the detection mechanism used in existing acoustic, ultrasound and RF localization systems in WSN.
\vspace{1mm}
\newline
\noindent
\textbf{Non Cross-correlation:} 
Active Bat \cite{Harter1999:Bat}, Cricket \cite{Priyanthathesis2005:Cricket}, Medusa \cite{Savvides2001:ALHOS} and SpiderBat \cite{Spiderbat:Oberholzer} are ultrasound positioning systems.
Range measurements are performed by calculating the TDOA between two synchronously sent RF and ultrasonic pulses at the receiver.
The ranging pulse is a single frequency ($40$\,kHz) sinusoidal and its arrival is detected by triggering an interrupt pin of the microcontroller when its leading edge exceeds a preset threshold.
Due to the functional simplicity, low-power microcontrollers (Atmega/MSP430 series) used in these platforms are efficient in managing the on-board processing.
Kusy et al. in \cite{interferometric:Kusy} introduced radio inferferometry to design a low-cost RF-based positioning system on the Mica2 platform.
This method measures the relative phase offsets of the interference field (created by two nodes transmitting RF pulses at slightly different frequencies) at different locations to obtain the position estimate of the transmitters.
However, these techniques are not robust against multipath characteristics, and so, no results have been published for complex cluttered environments.
\vspace{1mm}
\newline
\noindent
\textbf{Cross-correlation:}
The system proposed by Kushwaha et al. in \cite{Kushwaha2005}, Hazas et al. in \cite{Hazas2006:USBroadband}, AENSBox \cite{Girod2006:AENSBox}, BeepBeep \cite{Chunyi2007:BeepBeep} and TWEET \cite{Misra2011:tweet} are existing acoustic broadband systems.
Despite their difference in signal design, synchronization schemes and methods to improve the received SNR, they share a common detection mechanisms:cross-correlation.
These systems have been reported to withstand considerable channel multipath and environmental noise, and so, benefit in providing reliable and precise distance estimates for long coverage range. 
However, the capability of these systems have been upgraded by using DSP/smart phones that typically consume higher power and resources.  
\newline
\indent
The theory of sparse representation \cite{donoho:l1} helps to efficiently embed information without much loss (which serves the purpose of storage and transmission) followed by its recovery from an underdetermined system.
Although, we follow a similar approach as Wright et al. \cite{Wright2009:face} in face recognition, the scope of our problem is completely different.
We design a new dictionary, specifically, for cross-correlation based detection and ranging, as opposed to feature extraction for face classification.
\newline
\indent
Previous work by Whitehouse et al. in~\cite{whitehouse2002:calamari} and Sallai et at. in~\cite{Sallai04acousticranging} on acoustic ranging in resource constrained sensor networks (using MICA platform) categorically state that the limited availability of RAM was the most serious constraint in their system implementation.
The ranging results reported by \cite{Sallai04acousticranging} have an average error of $8.18$\,cm over a distance of $1$-$9$\,m by repeating the ranging signal $16$ times, which results in significant runtime and energy overhead.
Using \emph{StructS-XCorr}, our acoustic ranging system was able to confront this problem, and also, was able to provide similar performance (mean error of~$< 10$\,cm over $1$-$10$\,m) with fewer samples.
\newline
\indent
\emph{StructS-XCorr} approach has \emph{several merits}. 
\emph{First}, it provides a simple dimensionality reduction mechanism (that can be implementable on a typical WSN node) as a viable alternative to the computationally intensive cross-correlation function. 
\emph{Second}, it requires processing a significantly smaller datasets (proportional to the logarithmic count of the acquired signal samples) to obtain accuracies comparable to cross-correlation (the state-of-the-art detection technique).
At the local device end, the simplicity of this operation translates into appreciable resouce savings.
\emph{Finally}, it is independent of the physical signal (radio/acoustic) and medium (air/water), and is therefore a versatile framework for wide range of uses.
However, the centralized processing framework is the primary \emph{drawback}.
Towards this end, we \emph{argue} that it is a reasonable trade-off for achieving the performance of cross-correlation on mote-class devices.
We envision that, besides having an impact on current location sensing systems, it would create a new drive for WSN applications\cite{Misra2014:sparsegps} where the requirement for reliable location information on constrained network embedded sensing elements hold more importance than centralized computation.

\vspace{-2mm}
\section{Conclusion and Discussion} \label{sec:conclusion}
We presented a new information processing approach for range estimation: cross-correlation via sparse representation and structured sparsity.
We showed that exploiting structure of sparsity is critical for high-performance signal processing operations of high-dimensional data such as cross-correlation.
The sparsity of the underlying signal in our proposed correlation domain aids in the recovery mechanism to obtain reliable range estimates.
The main idea was to use a Hankel matrix with the time-shifted reference signal as the dictionary that leads to sparser representation than processing in other domains.
The design of the correlation dictionary and information recovery using structured sparsity are the main contributions of this work, which allows for ToA estimation with or without compressed sensing.
We designed its theoretical framework and validated its working through empirical system tests and characterization studies.
Considering the implementation simplicity in the acoustic domain, we developed an end-to-end acoustic ranging system using COTS sensor platforms to verify our hypothesis.
\newline
\indent
Our work in this paper is guided by the current hardware limitations of low-cost and low-power sensor platforms.
We believe that the key observations and principles derived here will find their application\cite{prasant_mining:2010,Misra2014:sparsegps} in location sensing systems that have constrained hardware resources to handle the bulk of data processing.

\vspace{-2mm}
\bibliographystyle{IEEEtran}
\bibliography{tmc-sparseXCORR}
\vspace{-13mm}
\begin{IEEEbiography}[{\includegraphics[width=1in,height=1.25in,clip,keepaspectratio]{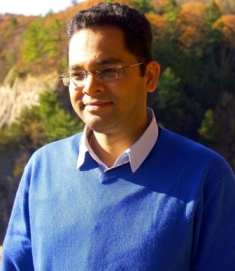}}]{Prasant Misra}
(S'08--M'12--SM'15) is a scientist at the TCS Innovation Labs (Bangalore).  
He received his Ph.D. from the University of New South Wales, Sydney in $2012$. 
He performs research in low-power and energy-efficient sensing/signal processing/wireless communication with a focus on system design for CPS/IoT.
He has many years of experience in research/technology development. He has worked in different roles and capabilities for: Keane Inc. (a unit of NTT Data Corporation), CSIRO, SICS Swedish ICT and RBCCPS@IISc Bangalore; the outcomes of which have either resulted in commercial/open-source solutions or publications in premier sensornet/IoT forums such as ACM/IEEE IPSN and ACM TOSN.
His professional contributions have been recognized by numerous awards, of which it is noteworthy to mention the ERCIM Alain Bensoussan, Marie Curie fellowship ($2012$) and the AusAID (Australian Government) Australia Leadership Awards scholarship ($2008$).
Currently, he  the secretary of IEEE Computer Society (Bangalore Section) and an Associate Technical Editor of IEEE Communication Magazine.
\end{IEEEbiography}

\begin{IEEEbiography}[{\includegraphics[width=1in,height=1.25in,clip,keepaspectratio]{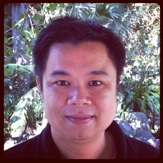}}]{Wen Hu} is a senior lecturer at School of Computer Science and Engineering, the University of New South Wales (UNSW). Much of his research career has focused on the novel applications, low-power communications, security and compressive sensing in sensor network systems and Internet of Things (IoT). Hu published regularly in the top rated sensor network and mobile computing venues such as ACM/IEEE IPSN, ACM SenSys, ACM Transactions on Sensor Networks (TOSN), Proceedings of the IEEE, and Ad-hoc Networks.
\newline\indent
Dr. Hu was a principal research scientist and research project leader at
CSIRO Digital Productivity Flagship, and received his Ph.D from the
UNSW. He is a recipient of prestigious CSIRO Office of Chief Executive
(OCE) Julius Career Award (2012 - 2015) and multiple CSIRO OCE
postdoctoral grants.
\newline\indent
Dr. Hu is a senior member of ACM and IEEE, and serves on the organising
and program committees of networking conferences including ACM/IEEE
IPSN, ACM SenSys, ACM MobiSys, IEEE ICDCS, IEEE LCN, IEEE ICC, IEEE
WCNC, IEEE DCOSS, IEEE GlobeCom, IEEE PIMRC, and IEEE VTC.
\end{IEEEbiography}
\vspace{-13mm}
\begin{IEEEbiography}[{\includegraphics[width=1in,height=1.25in,clip,keepaspectratio]{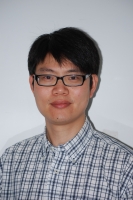}}]{Mingrui Yang}
is a Senior Research Associate in the Department of Radiology at the Case Western Reserve University, Cleveland, OH. He was an OCE Postdoctoral Research Fellow in Digital Productivity Flagship of the Commonwealth Scientific and Industrial Research Organization (CSIRO), Australia. He received his Ph.D. in Mathematics from the University of South Carolina, USA in 2011. He is interested in interdisciplinary researches in compressive sensing, sparse approximation, signal/image processing, greedy algorithms and nonlinear approximation, and their applications in sensor networks, hyperspectral imaging and magnetic resonance imaging. He has published in high quality journals and conferences in mathematics and computer science, such as Advances in Computational Mathematics, IEEE Transactions on Signal Processing, IEEE Transactions on Mobile Computing, ACM/IEEE IPSN, and ACM SenSys. He is a member of IEEE and ISMRM.
\end{IEEEbiography}
\vspace{-13mm}
\begin{IEEEbiography}[{\includegraphics[width=1in,height=1.25in,clip,keepaspectratio]{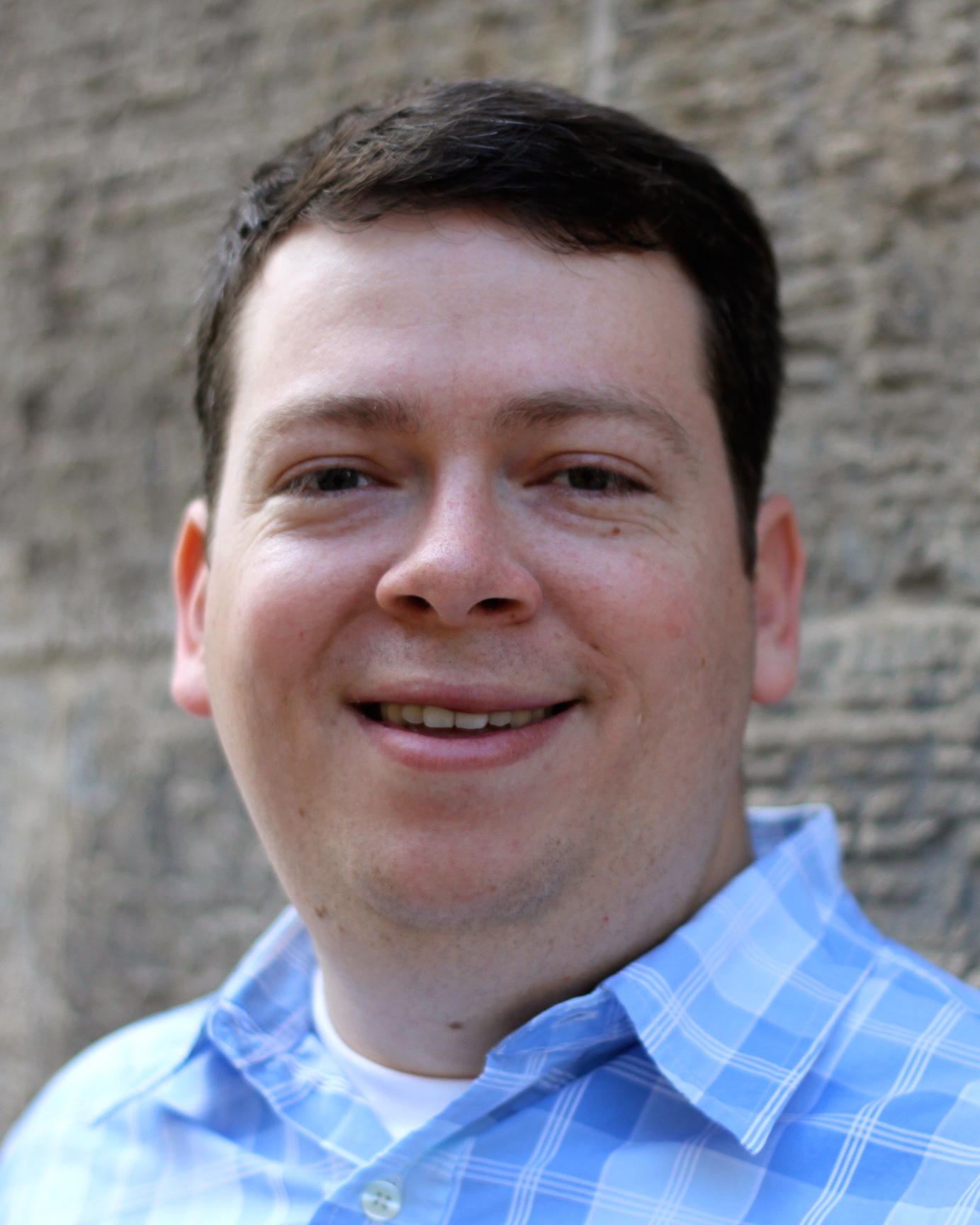}}]{Marco F. Duarte}
(S'99--M'09--SM'14) received the B.Sc.\ degree in computer engineering (with distinction) and the M.Sc.\ degree in electrical engineering from the University of Wisconsin--Madison in 2002 and 2004, respectively, and the Ph.D.\ degree in electrical engineering from Rice University, Houston, TX, in 2009. He was an NSF/IPAM Mathematical Sciences Postdoctoral Research Fellow in the Program of Applied and Computational Mathematics at Princeton University, Princeton, NJ, from 2009 to 2010, and in the Department of Computer Science at Duke University, Durham, NC, from 2010 to 2011. He is currently an Assistant Professor in the Department of Electrical and Computer Engineering at the University of Massachusetts, Amherst, MA. His research interests include machine learning, compressed sensing, sensor networks, and computational imaging.
\newline
\indent
Dr. Duarte received the Presidential Fellowship and the Texas Instruments Distinguished Fellowship in 2004 and the Hershel M. Rich Invention Award in 2007, all from Rice University. He is also a member of Tau Beta Pi.
\end{IEEEbiography}
\vspace{-13mm}
\begin{IEEEbiography}[{\includegraphics[width=1in,height=1.25in,clip,keepaspectratio]{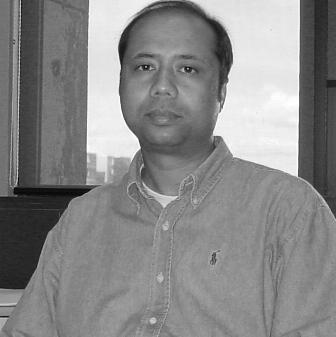}}]{Sanjay Jha} is a Professor and Head of the Network Group at the School of Computer Science and Engineering at the University of New South Wales.   His research activities cover a wide range of topics in networking including Network and Systems Security,  Wireless Sensor Networks, Adhoc/Community wireless networks, Resilience and Multicasting in IP Networks.  Sanjay has published over $160$ articles in high quality journals and conferences. He is the principal author of the book Engineering Internet QoS and a co-editor of the book Wireless Sensor Networks: A Systems Perspective. He served as an associate editor of the IEEE Transactions on Mobile Computing (TMC) was on the editorial board of the ACM Computer Communication Review (CCR).
\end{IEEEbiography}

\end{document}